\documentclass[reprint, amsmath, amssymb, aps, pra, showkeys]{revtex4-2}
\usepackage{xfrac}
\usepackage{amsmath}
\usepackage{bmpsize}
\usepackage{graphicx}
\usepackage{svg}
\usepackage{dcolumn}
\usepackage{bm}
\usepackage{blindtext}
\usepackage{braket}
\usepackage{times} 
\usepackage{lmodern} % Font types
\usepackage{anyfontsize} % Font types

\usepackage{bibunits}
\usepackage{nameref}
\usepackage[colorlinks=true, allcolors=blue]{hyperref}

\hypersetup{
    colorlinks=true,
    linkcolor=blue,
    filecolor=magenta,      
    urlcolor=cyan,
    citecolor=blue,
    pdftitle={Nasser et. al., On-Demand},
    pdfpagemode=FullScreen}
\usepackage[utf8]{inputenc} % Ensure UTF-8 encoding is enabled
\DeclareUnicodeCharacter{A76C}{,} % Replace A76C (ꎬ) with a comma or any other symbol
\usepackage[T1]{fontenc}
\usepackage{placeins}  % Add to preamble if not already included
\FloatBarrier  % Forces all previous floats to be placed

\begin{document}

\preprint{APS/123-QED}

\title{Memory-assisted multimode microwave-to-optical transduction}

\author{Ujjwal Gautam$^{1,2}$}
\author{Nasser Gohari Kamel$^{1,2}$}
\author{Sourabh Kumar$^{1,2}$}
\author{Daniel Oblak$^{1,2}$}
\email{doblak@ucalgary.ca}

\affiliation{$^{1}$Institute for Quantum Science and Technology, University of Calgary, Calgary, AB, Canada T2N 1N4}
\affiliation{$^{2}$Department of Physics and Astronomy, University of Calgary, Calgary, AB, Canada T2N 1N4}
\homepage{https://qcloudlab.com/}

\date{\today}

%%%%%%%%%%%%%%%%%%%%%%%%%%%%%%%%%%%%%%%%%%%%%%%%%%%%%%%%%%%%%%%%%%
\begin{abstract}
Microwave-to-optical quantum transducers will enable coherent interconnection between distant superconducting quantum devices. Ongoing explorations with several platforms have shown promising results at single-photon levels. However, in all these demonstrations, elimination of noise due to the concurrence of the weak transduced signal with intense pump pulses remains a challenge, requiring high suppression filtering setups. A memory-assisted transducer, on the other hand, offers a versatile approach that not only mitigates the noise but also enables the on-demand retrieval of the transduced signal. Here, we integrate a quantum memory protocol with transduction in a three-level atomic system to demonstrate on-demand retrieval of transduced signals. Due to the zero-first-order Zeeman transitions at zero magnetic fields, providing long optical and spin coherence times, and GHz range hyperfine splitting, we use a low-doping concentration $^{171}{\rm Yb}^{3+}$:${\rm Y}_2{\rm SiO}_5$ crystal at 30\,mK temperature. We achieve on-demand transduction assisted by memory with $0.4\ (\text{and }0.3)$ noise photons in the detection window at a storage duration of $460\ (\text{and }620) \, \mu \textrm{s}$. To demonstrate the coherent nature of the protocol, we show interference patterns resulting from transduced signals due to varying phase or frequency of the input microwave pulses. Further, multimode transduction capacity is demonstrated, utilizing the spin and optical inhomogeneous broadening. The on-demand capability of the protocol allows synchronizing qubits in a quantum repeater protocol, while multimode capacity increases the entanglement generation rate. To the best of our knowledge, this is the first demonstration of an on-demand microwave-to-optical transducer assisted by memory.
\end{abstract}

\maketitle

%%%%%%%%%%%%%%%%%%%%%%%%%%%%%%%%%%%%%%%%%%%%%%%%%%%%%%%%%%%%%%%%%%%%%%%%%%%%
\section{Main} \label{sec:Main}
\begin{figure*}[htbp]
    \centering  \includegraphics[width=1\linewidth]{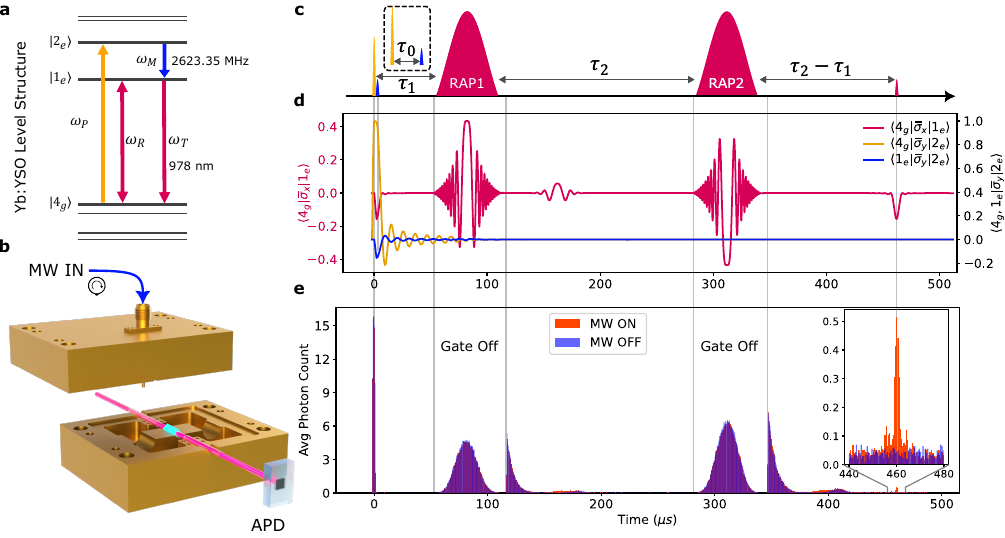}
    \caption{\textbf{Protocol Sequence.} \textbf{a,} A three-level system in Yb:YSO with excited state spin transition showing respective colour-coded transitions used in \textbf{c}. \textbf{b,} Experimental setup depicting Yb:YSO crystal (cyan) kept at the centre loop of a 3D loop-gap resonator. SMA antenna pin couples MW in and out via an isolator. A pair of GRIN lenses is used for optical coupling through the crystal. \textbf{c,} Schematic of the pulse sequence. Intense pump pulse (yellow) is followed by a weak coherent MW pulse (blue) after a time $\tau_0$ (inset). RAP1 and RAP2 pulses (red) of duration $\tau_R$ are applied after $\tau_1$ from the pump pulse and after $\tau_2$ from RAP1, respectively. Transduced optical signal (red) is emitted at $\tau_2- \tau_1$ from RAP2. Total storage time is $2(\tau_2+ \tau_R)$. \textbf{d,} Simulated evolution of relevant coherences excited by the corresponding colour pulses. Coherence in $\langle 4_g|\overline\sigma_x |1_e\rangle$ is a result of a two-pulse excitation which is then rephased with RAP1 and RAP2. \textbf{e,} \textcolor{black}{Mean photon counts per 500\,ns bin, averaged over 200 experimental cycles.} The tall bar at $t=0$ is the leaked part of the pump pulse. Leaked RAP1 and RAP2 are seen in the gate OFF windows. Decaying counts after the RAP pulses are the corresponding free-induction decay. The transduced echo is visible when the MW is ON. (Inset) Zoom-in view around the transduced echo showing signal versus the noise floor counts.} \label{fig:protocol}
\end{figure*}

Reliable and fast distribution of entanglement between quantum devices is a hallmark of future quantum networks~\cite{Kimble_2008, hu-2021}. With long-distance communication relying on flying optical qubits while microwave (MW) qubits used for local information processing~\cite{acharya-2024} or sensing~\cite{hecht-2025, pirandola-2018}, a microwave-to-optical (M2O) transducer will be essential to bridge these disparate modalities~\cite{xie-2025, caleffi-2025, kumar-2019, lauk-2020, han-2021}. M2O transducers also find near-term application as interconnects between superconducting quantum circuits (SQCs) in a distributed quantum computing architecture~\cite{cirac-1999}, to overcome scalability issues that currently limit these systems~\cite{megrant-2025}.
An M2O transducer must meet several challenging requirements, such as, operating at $\ll\,1$ noise photons per qubit mode for quantum applications, while achieving high efficiency and large bandwidth, for faster operation~\cite{lauk-2020, han-2021, zeuthen-2020}. Moreover, a large multiplexing capacity will enable simultaneous transduction of several modes~\cite{smith-2023}, further increasing the qubit transfer rate while storage and on-demand retrieval of the transduced optical qubit~\cite{kurokawa-2022} are paramount for synchronization in a path-erasing entanglement scheme~\cite{zeuthen-2020}. Finally, since magnetic (B) fields induce noise in SQCs~\cite{kakuyanagi-2007}, an M2O transducer should ideally operate at $B=0$. \par 

Several platforms have demonstrated M2O transduction, including electro-optics~\cite{sahu-2022}, opto-mechanics~\cite{higginbotham-2018, weaver-2023}, and atomic ensembles~\cite{tu-2022, xie-2025}, and some integration with SQCs for coherent control and readout~\cite{mirhosseini-2020, delaney-2022}. As an alternative to direct M2O transduction, MW-optical entangled photon-pairs have also been generated~\cite{meesala-2024, sahu-2023}. Since, M2O transduction relies on a strong pump field to achieve significant non-linear response and energy-conservation~\cite{han-2021}, suppression of noise photons stemming from the pump field remains a significant challenge, typically necessitating lossy filtering setups~\cite{rochman-2023}. Memory-assisted transduction~\cite{kurokawa-2022}, on the other hand, delays the transduced optical signals with respect to the pump field and associated noise~\cite{xie-2025}. 
In addition, memory-assisted transduction is less sensitive to the temporal shape/mode of signal photons and, thus, more practical for qubit encoding, e.g., as time-bins \cite{singh2025photonic}.
The quest for systems that can deliver on all transduction figures-of-merit together with memory capability, continues to be of high priority. \par

One approach to quantum transduction is based on the coupling of optical and MW transitions in ensembles of rare-earth ions (REIs), e.g., Er$^{3+}$~\cite{king-2024, rochman-2023, chaneliere-2024, probst-2013} and Yb$^{3+}$~\cite{xie-2025} doped in various host crystals.
Although Er$^{3+}$ conveniently possesses an optical transition in the telecom-band ($\sim 1530~\textrm{nm}$), the application of a B-field is required (even isotopes) to split its Zeeman levels for the MW spin transition, complicating integration with SQCs~\cite{king-2024}. 
A transduction efficiency of $10^{-5}$ was shown for a 10\,ppm doped Er:YSO crystal placed in an optical cavity and a 3D loop-gap resonator (LGR)~\cite{ball-2018} with a $146\,\textrm{mT}$ external B-field~\cite{fernandez-gonzalvo-2019}. 
In contrast, $^{171}$Yb$^{3+}$ inherently possesses hyperfine and zero first-order Zeeman (ZEFOZ) transitions at zero B-field~\cite{tiranov-2018}. The 1\% efficiency demonstrated in 340\,ppm doped Yb:YVO crystal with a planar MW resonator and a weak optical cavity~\cite{xie-2025}, has served as a benchmark for REI-based transduction.\par

In this work, the MW spin and optical electron transitions in a 5\,ppm-doped Yb:YSO crystal \textcolor{black}{with no applied B-field} are employed to demonstrate on-demand, memory-assisted M2O transduction of up to ten multiplexed modes. 
The unique memory functionality not only enables on-demand retrieval of the transduced signal but also eliminates the need for lossy spectral filtering of the pump. We report an internal transduction efficiency of $4\times10^{-8}$
for a 460\,$\mu \mathrm{s}$ storage time, without the aid of an optical cavity. 
\textcolor{black}{The storage-time dependent noise-floor in the output optical signal, dominated by free-induction decay (FID) from the co-propagating frequency-chirped memory control-pulses, is measured at 0.4 (and 0.3) photons for 460\,$\mu\mathrm{s}$ (and 620\,$\mu\mathrm{s}$) storage time.} 
We verify preservation of coherence of the transduced optical signals through observing interference between two MW pulses with varying phase- and frequency difference and between the transduced optical signal and a reference local oscillator.
Finally, we demonstrate spectral and temporal multiplexing enabled by the broad inhomogeneous spin and optical linewidths of the Yb:YSO system.\par

%%%%%%%%%%%%%%%%%%%%%%%%%%%%%%%%%%%%%%%%%%%%%%%%%%%%%%%%%%%%%%%%%%%%%%%%%%%%%%%%%%%%%%
\section{Results}\label{sec:Results}

\subsection{Memory-assisted M2O Transduction Protocol}
\label{sec:protocol}
Generally M2O transduction converts a MW signal at frequency $\omega_M$ to a transduced optical signal at frequency $\omega_T$ using an optical pump field at frequency $\omega_P$, while maintaining energy conservation, $\omega_T =\omega_P \pm \omega_M$. We choose a three-level sub-system of Yb:YSO at zero B-field (Fig.\,\ref{fig:protocol}\textbf{a}) and drive the excited state spin transition $|1_e\rangle \leftrightarrow|2_e\rangle$ (blue) at $\omega_M = 2623.35\,\textrm{MHz}$ with an LGR at 2625.3\,MHz and 4\,MHz linewidth (Fig.\,\ref{fig:protocol}\textbf{b}). The optical transitions $|4_g\rangle \leftrightarrow|2_e\rangle$ and $|4_g\rangle \leftrightarrow|1_e\rangle$ at $\omega_P$ and $\omega_T \ (= \omega_P - \omega_M)$ respectively, are coupled by free-space optical fields, and the weak transduced signal is detected on an avalanche photodiode (APD). Although the optical transitions are ZEFOZ 
% $|4_g\rangle \leftrightarrow |1_e\rangle$ is a ZEFOZ transition 
at zero magnetic field, we do not actively compensate for residual environmental magnetic field, which is likely on the order of a few $\mu\textrm{T}$~\cite{nicolas-2023}.

The memory-assisted M2O transduction sequence (shown in Fig.\,\ref{fig:protocol}\textbf{c}) is an extension from a two-level QM protocol in which rephasing of a collective atomic excitation is achieved on-demand using a pair of Rapid Adiabatic Passage (RAP) pulses~\cite{kamel-2025, osullivan-2022, pascual-winter-2013}. To understand the atomic rephasing, it is informative to visualize the evolution of relevant collective ensemble coherences $\langle \overline\sigma_{x,y} \rangle$ (SI~Sec.~II~\cite{gautam2025spp}) during the transduction sequence as shown in Fig.\,\ref{fig:protocol}\textbf{d} (SI~Fig.~S5~\cite{gautam2025spp} for all the coherence components). An intense pump pulse of duration $\tau_P$ at frequency $\omega_P$, initiates $\langle 4_g|\overline\sigma_y |2_e\rangle$ optical coherence at $t=0$. A weak MW pulse applied after a delay $\tau_0$ interacts with the excited state population and generates $\langle 2_e|\overline\sigma_y |1_e\rangle$ spin coherence. The cumulative effect of the two-pulse excitation results in $\langle 4_g|\overline\sigma_x |1_e\rangle$ optical coherence on the transduced signal transition, which dephases and averages out to zero over the ensemble (Fig.\,\ref{fig:protocol}\textbf{d}). Similar to that for an optical QM in a two-level system~\cite{kamel-2025}, this coherence is rephased by a pair of RAP pulses centred at frequency $\omega_R \sim \omega_T$. Both the RAP pulses, are identical, amplitude- and phase- modulated pulses, with duration $\tau_R$ and linear frequency chirp bandwidth $\Delta_{\rm{ch}}$ corresponding to a chirp rate of $R_{\rm{ch}}=\Delta_{\rm{ch}}/\tau_R$. For optimal rephasing, the RAP pulses must satisfy the adiabaticity condition $\Omega _R^2/R_{\rm{ch}} \gg 1$, where $\Omega_R$ is the resonant Rabi frequency of the RAP pulses. After $\tau_2-\tau_1$ duration from RAP2, the collective coherence, $\langle 4_g|\overline\sigma_x |1_e\rangle$ rephases and emits at frequency $\omega_T$ in the optical domain. Here, the temporal mark and the temporal profile of the transduced signals are based on the initial pump pulse, thereby defining the storage time as $\tau_s=2(\tau_2+\tau_R)$. 
Note that the temporal separation between the transduced signal and the control pulses avoids the need for spectral filtering.\par

Fig.\,\ref{fig:protocol}\textbf{e} is an experimental result of the protocol for which the unabsorbed pump-pulse ($\tau_P=4\,\mu\textrm{s}$) at $t=0$, results in the initial peak. The MW pulse of duration $\tau_M=2\,\mu\textrm{s}$ is applied right after the pump pulse ($\tau_0=0\,\mu\textrm{s}$) followed by the RAP pulses with $\tau_R=60\,\mu$s and $R_{\text{ch}}=2\pi\times 25\,\mathrm{MHz/ms}$ are applied with $\tau_1=50\,\mu\textrm{s}$, and $\tau_2=170\,\mu\textrm{s}$. 
This results in the emission of a transduced signal at $t=460\,\mu\textrm{s}$ following the pump pulse. During the intense RAP pulses, the APD is gated OFF to avoid detector damage and saturation. The exponentially decaying counts after the RAP pulses are coherent FID emission~\cite{yu-2010} (Sec.~\ref{sec:noise}).\par

\begin{figure}[htbp]
    \centering  \includegraphics[width=0.85\linewidth]{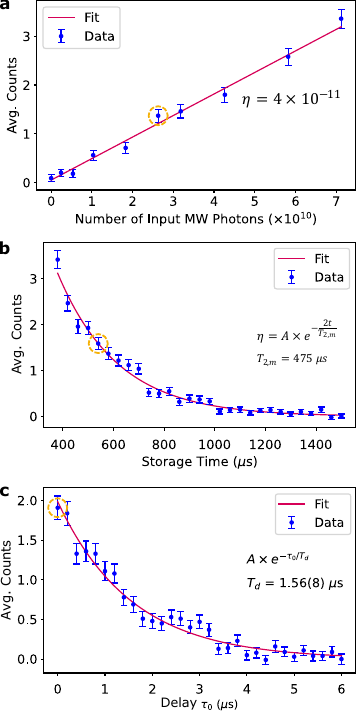}
    \caption{\textbf{Characterizations.} Sweep of various parameters showing transduced signal counts in a $4\,\mu \mathrm{s}$ detection bin. The yellow ring depicts the parameters used across experiments. \textbf{a,} Sweep of number of input MW photons per pulse resulting in a linear rise in transduced counts. The slope of the linear fit gives an overall photon conversion efficiency of $4\times 10^{-11}$. \textbf{b,} Sweep of storage duration vs transduced signal counts adjusted for the noise floor. An exponential decay fit function of the form $A\times\exp{-2t/T_{2,m}}$, resulted in $T_{2,m} = 475\,\mu \mathrm{s}$ as the memory coherence time. \textbf{c,} Sweep of pump-MW delay $\tau_0$, resulted in a phenomenological decay with time constant of $T_d = 1.56\pm 0.08\, \mu \mathrm{s}$.}  \label{fig:Characterizations}
\end{figure}

\begin{figure*}[ht]
    \centering  \includegraphics[width=1\linewidth]{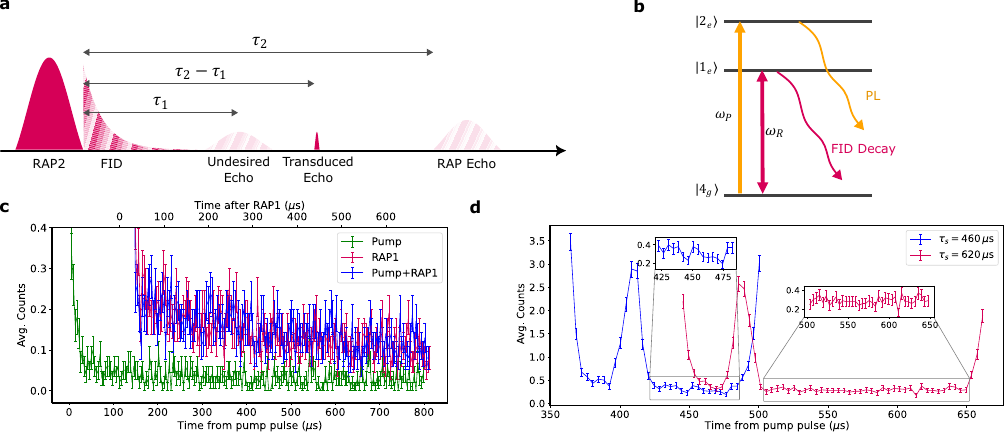}
    \caption{\textbf{Noise and Transduced echo placement.} \textbf{a,} Schematic of optical pulse (solid) and noise regions (shaded) around transduced echo. \textbf{b,} Level diagram with sources of noise from pump (yellow) and RAP (red). \textbf{c,} Noise photon counts due to pump pulse and RAP1. Each data point is photon counts in $4\, \mu\mathrm{s}$ integration window with bottom time axis referenced to the pump, while top time axis referenced to tail of RAP1. Green data points are noise due to only pump (Photoluminescence), red due to only RAP1 (Free-Induction Decay), and blue due to both pump and RAP1. \textbf{d,} Average photon counts around the expected transduced signal region for two storage durations with pump and both the RAP pulses ON while keeping the MW pulse OFF. The valley (box) between the peak and the higher counts towards the end is the region where the echo needs to be placed.}  \label{fig:Noise}
\end{figure*}

\begin{figure*}[ht]
    \centering  \includegraphics[width=1\linewidth]{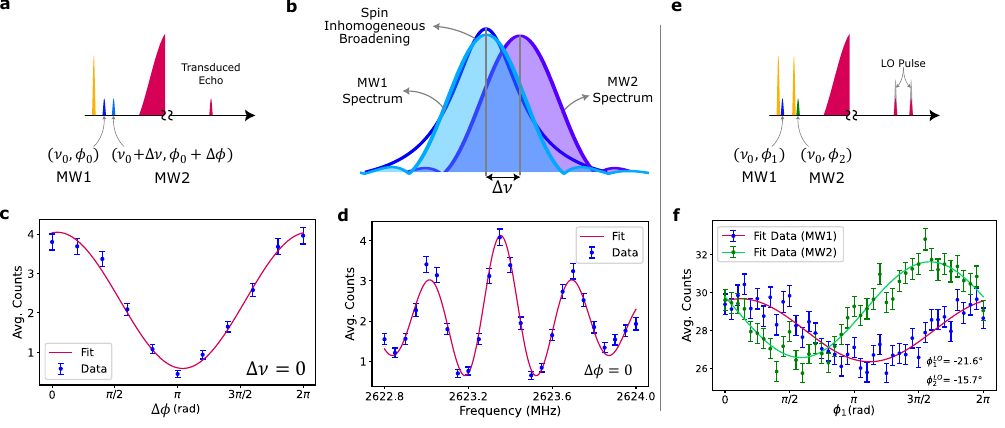}
    \caption{\textbf{Interference.} \textbf{a,} Schematic of the pulse sequence consisting of a pump pulse (yellow) followed by two MW pulses (blue) with detuning ($\Delta\nu$) and phase difference ($\Delta\phi$) resulting in a single transduced echo. There is no delay between the pulses. Gaps are shown to highlight the individual pulses. \textbf{b,} Schematic of the overlap between the Fourier spectrum of the individual pulses and the spin inhomogeneous broadening. \textbf{c,} \textcolor{black}{Sweep of the phase difference ($\Delta\phi$) showing interference pattern after adjusting for noise-level. Data fit with a sinusoidal interference function $I_1+I_2 + 2\sqrt{I_1I_2} \cos(\Delta\phi+\phi_{0}))$, where $I_1 (1.93)$ and $I_2 (0.39)$ correspond to the amplitudes of the transduced signals due to MW1 and MW2, respectively and $\phi_{0}\ (-7.3^\circ)$ is a fit parameter.} \textbf{d,} \textcolor{black}{Sweep of Frequency of MW2 for a set phase difference $(\Delta\phi \ =0)$ showing interference pattern after adjusting for noise level. Data is fitted with a Lorentzian modulated interference function to account for the spin inhomogeneous broadening: $I_1 + I_2 \alpha+2\sqrt{I_1I_2\alpha}\cos(k \Delta\nu)$, where $\alpha = L(\nu, \nu_0)$ is the Lorentzian function at frequency $\nu$ centered at $\nu_0$. $I_1\ (1.83)$ and $I_2\ (0.46)$ correspond to the amplitudes of the transduced signals due to MW1 and MW2, respectively. Here, $k\ (0.0181)$ is a fit parameter which denotes the wavenumber of the optical fields.} \textbf{e,} Schematic of pulse sequence showing a pair of pump and MW pulses being transduced to a pair of optical signals. Local oscillator (LO) pulses are synchronized to arrive with the transduced signals at the APD. \textbf{f,} Sweep of phase of MW1, $\phi_1$ and MW2 with $\phi_2 = \phi_1 + \pi/2$, showing interference pattern with LO pulses of constant phase. \textcolor{black}{Data is fitted with $A(1+V\cos{(\phi_{1,2} + \phi_{1,2}^{\textrm{LO}})})+C$, where $V$ ($\sim82\%$ for both MW1 and MW2) is the visibility of the interference, $\phi_{1,2}^{\textrm{LO}}$ are the phases of the LO pulses for each mode and $C\ (26)$ is fit parameter for the noise level due to leak light.} Fit values are mentioned in the parentheses.} \label{fig:interference}
\end{figure*}

\subsection{Protocol Characterization}

\subsubsection{Input MW Photons and Efficiency}
The transduction efficiency in Fig.~\ref{fig:Characterizations}\textbf{a} is characterized under optimal pulse parameters for $\tau_s=460\,\mu\mathrm{s}$. The input MW photons are varied in weak-power regime to ensure that the spin transition remains unsaturated. After accounting for losses and inefficiencies of the experimental setup, the internal memory-assisted transduction efficiency is $4 \times 10^{-8}$. \textcolor{black}{A detailed discussion of efficiency budget is presented in Sec.~\ref{subsec:efficiency_budget}.}\par

\subsubsection{Memory Coherence Time}
The collective coherence created on the $\ket{4_g} \leftrightarrow \ket{1_e}$ transition decoheres due to the finite optical coherence time, thus degrading the transduction efficiency at longer storage durations. To quantify this effect, we vary $\tau_2$ in $\tau_s$ in Fig.~\ref{fig:Characterizations}\textbf{b}, and baseline-correct the counts to subtract the noise floor. The exponential fit yields memory coherence time $T_{2,m} = 475\,\mu\mathrm{s}$\textcolor{black}{, well within the optical coherence time of the transition $T_{2,o} = 801\,\mu\mathrm{s}$ (SI~Fig.~S7\textbf{f}~\cite{gautam2025spp}).}\par

\subsubsection{Pump-MW delay}
Following the pump pulse, each atom accumulates a phase determined by both its optical and spin detunings. Over time, the dephasing on the spin transition leads to phase scrambling of the atomic coherence. As a result, any delay in transferring the collective coherence from the pump transition reduces the final coherence on the RAP transition, confirmed from simulations (SI~Fig.~S6~\cite{gautam2025spp}).
Fig.~\ref{fig:Characterizations}\textbf{c} presents the experimental data fitted with exponential decay, yielding a time constant of $T_d = 1.56 \pm 0.08\,\mu\mathrm{s}\sim 641\,\mathrm{kHz}$, matching \textcolor{black}{the spin-inhomogeneous broadening} $\Gamma_s$ (SI~Fig.~S7\textbf{c}~\cite{gautam2025spp}).\par

\subsection{Noise Characterization}
\label{sec:noise}
The protocol provides some freedom to choose the recall time of the transduced signal. With the goal of achieving the optimal signal-to-noise ratio, the transduced signal should avoid overlap with noise expected at certain times in the protocol, indicated by the shaded pulses in Fig.\,\ref{fig:Noise}\textbf{a}. We, thus, characterize the noise, taken as the integrated optical photon counts within a window of 4\,$\mu\mathrm{s}$, corresponding to the duration of a potential transduced signal, at different times in the sequence.

One source of noise is caused by the excitation of population by the pump pulse from the ground state, $\ket{4_g}$, to the excited state, $\ket{2_e}$, which decays back to the ground state via photoluminescence (PL), as illustrated in Fig.\,\ref{fig:Noise}\textbf{b}. As this PL frequency is distinct from that of the transduced signal, spectral filtering is possible. Engaging only the pump pulse, we see from Fig.\,\ref{fig:Noise}\textbf{c}, that PL from pump pulse contributes only $\sim 0.02$ photons to the noise floor, comparable to the detector's dark count rate, rendering any spectral filtering to be of low priority.\par

Another more significant contribution to the noise counts arises from the FID due to the co-propagating RAP pulses~\cite{de-seze-2005, yu-2010, dajczgewand-2014}, as illustrated in Fig.\,\ref{fig:Noise}\textbf{b}. The FID could be avoided by a slight spatial mismatch between the RAP and pump pulses~\cite{ma-2021}, but this was not possible in our setup due to practical limitations. We measure the FID noise by applying only the first RAP pulse and find in Fig.\,\ref{fig:Noise}\textbf{c} that at $\tau_s =460\,\mu\mathrm{s}$, RAP1 contributes $\sim 0.15$ photons to the noise floor. Assuming FID counts from RAP1 and RAP2 have similar decay patterns, RAP2 will contribute $\sim 0.25$ photons after $\tau_2 - \tau_1 = 120\,\mu \mathrm{s}$ from the end of the RAP pulse.\par

By engaging the full optical pulse sequence, we can assess the total noise contribution at $\tau_s = 460\,\mu\mathrm{s}$ from the pump and both the RAP pulses with MW OFF is 0.4 photons (Fig.\,\ref{fig:Noise}\textbf{d}), matching the sum of the individual contributions. However, in this case we also observe an undesired echo (peaks at $\sim 420\,\mu\mathrm{s}$ and $\sim 480\,\mu\mathrm{s}$, for the two storage times, respectively), as well as an RAP echo \textcolor{black}{responsible for the increase of noise at the tail of the two traces of Fig.\,\ref{fig:Noise}\textbf{d}}. \textcolor{black}{Note that noise at the start of the two traces corresponds to FID from the RAP pulses as also visible in Fig.\,\ref{fig:Noise}\textbf{c}}. The undesired echo arises from a two-pulse photon-echo (2PPE)-like process between the pump and the RAP pulses due to the zeroth order of the phase modulator (PM) leaking light at the same \textcolor{black}{fixed} frequency, \textcolor{black}{resonant with a separate atomic sub-ensemble} (SI~Fig.~S2~\cite{gautam2025spp}). Such an undesired echo also arises when the MW is ON and the adiabaticity condition is weakly satisfied (orange bumps in Fig.\,\ref{fig:protocol}\textbf{e}). The RAP echo arises due to the 2PPE-like scheme between the RAP pulses, i.e., the RAP2 pulse acts as a rephasing pulse for RAP1.\par

Returning to the objective of avoiding noise, we conclude that the valley between the undesired echo and the RAP echo is the sweet spot for retrieving the transduced signals, with less stringent temporal placement and lower noise counts for longer $\tau_s$ (0.3 for $\tau_s = 620\,\mu \mathrm{s}$) (Fig.\,\ref{fig:Noise}\textbf{d}).\par

\subsection{Coherent M2O Conversion}
\label{CoherentM2O}

\textcolor{black}{To demonstrate coherent information mapping, we apply an optical pump pulse followed by two sequential MW pulses with either varying phase difference ($\Delta\phi$) or detuning ($\Delta\nu$) (Fig.\,\ref{fig:interference}\textbf{a}). 
Each MW pulse determines the axis of rotation of the spin ensemble (Bloch vector) on the Bloch sphere based on its phase and detuning from the centre of the spin inhomogeneous broadening (Fig.\,\ref{fig:interference}\textbf{b}). For instance, the first MW pulse with $\Delta \nu=0$ and $\Delta \phi=0$, rotates the Bloch vector about the $+x$-axis, whereas the second MW pulse with $\Delta \nu=0$ and $\Delta \phi=\pi$, rotates it about the $-x$-axis, effectively reversing the initial rotation (Fig.~\ref{fig:interference}\textbf{c}). In another instance, if the first MW pulse is the same i.e. rotates about the $+x$-axis ($\Delta \nu=0$ and $\Delta \phi=0$), while the second MW pulse has $\Delta \nu\ne0$ and $\Delta \phi=0$, the axis of rotation tilts in the x-z plane, effectively bringing (taking) the Bloch vector closer (away) to (from) the pole (Fig.~\ref{fig:interference}\textbf{d})~\cite{foot-2005}. See SI Sec.~IV\,A~\cite{gautam2025spp} for a detailed description on two-pulse ensemble interaction.} 

\textcolor{black}{The ensemble Bloch vector after the sequence of MW pulses is mapped onto the transduced optical field, result in interference patterns as shown in Figs.~\ref{fig:interference}\textbf{c} and \textbf{d}. The fitting function described in the caption of Fig.~\ref{fig:interference}\textbf{c} fits well with $I_1=1.93$ and $I_2=0.39 \sim I_1\times\exp{(-\tau_M/T_d)}=0.54$, close to the photon count expected due to $\tau_0=\tau_M$ for the second MW pulse (Fig.~\ref{fig:Characterizations}\textbf{c}). The observed visibility can be calculated as $V=2\sqrt{I_1I_2}/(I_1+I_2)=75\%$, while the expected visibility, if $I_2$ is adjusted for decay $I_2^*=I_2\exp{(\tau_M/T_d)}=1.39$ is $V_{\mathrm{exp}}=98.7\%$. In Fig.~\ref{fig:interference}\textbf{d}, the observed and expected visibility at $\Delta\nu=0$ are $V=80\%$ and $V_{\mathrm{exp}}=99.8\%$, respectively.
As $\Delta\nu$ varies, the modulation decays due to the Lorentzian shape of the spin inhomogeneous broadening. 
This interference is well described by a Lorentzian modulated cosine function, directly demonstrating that phase information encoded in the MW field is coherently preserved and transferred to the transduced optical mode. Such faithful mapping of phase coherence enables protocols such as quadrature phase-shift keying (QPSK) for quantum information processing~\cite{zhou2025kilometer}.}

An alternative method to show coherent information transfer is by using Homodyne interference of the transduced signal with an optical local oscillator (LO) pulse (Fig.\,\ref{fig:interference}\textbf{e}). To achieve high-visibility interference, the LO pulses must be precisely synchronized in time, matched in optical frequency, and have similar photon counts as the transduced pulses. \textcolor{black}{The lowest photon counts we could reach in LO pulses were almost same as the transduced optical counts but with comparatively large leak photon counts which shifted the base-level of the interference pattern.} In Fig.\,\ref{fig:interference}\textbf{f} the phases of the MW pulses were varied with $\phi_1$ varying from 0 to $2\pi$ and $\phi_2 = \phi_1 + \pi/2$. Sinusoidal fit to the data maintains the constant phase of the LO pulses, thereby proving the coherent phase transfer. Further data on interference is presented in SI~Fig.~S11~\cite{gautam2025spp}.\par

\subsection{Spectro-Temporal Multiplexing}
\label{sec:multiplexing}
\begin{figure*}[htbp]
    \centering  \includegraphics[width=1\linewidth]{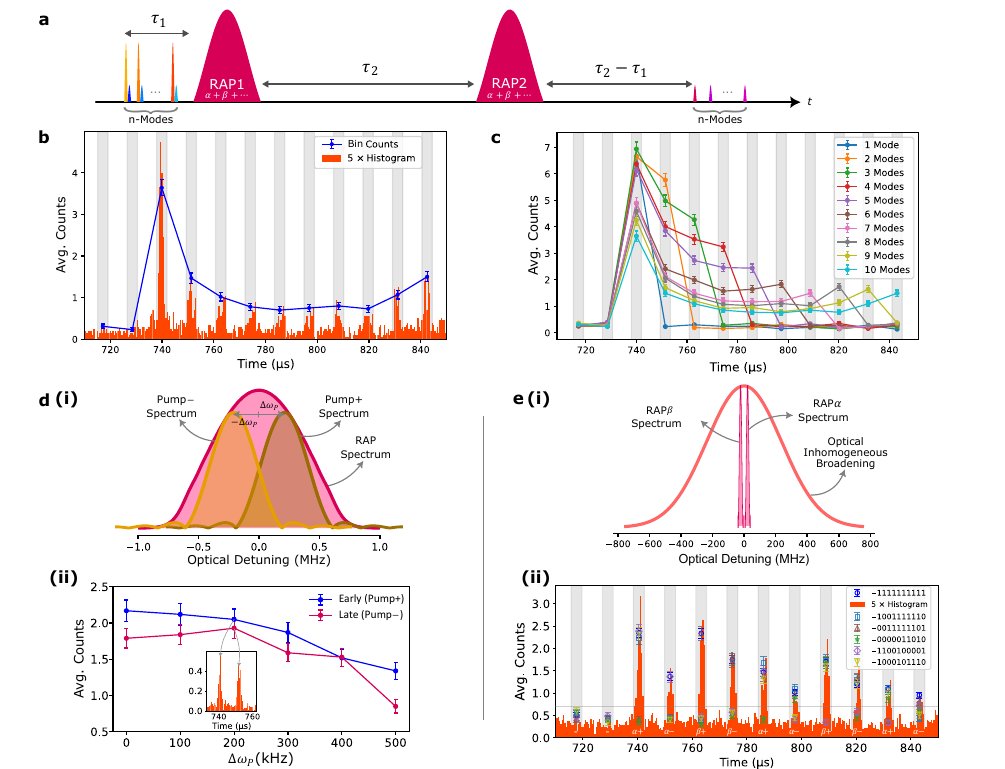}
    \caption{\textbf{Spectro-Temporal Multiplexing.} \textbf{a,} Schematic of the pulse sequence to store n-temporal modes and retrieve them in FIFO scheme using a pair of RAP pulses either at single or multiple frequencies. \textbf{b,} Experimental histogram data (orange) for 10-temporal modes. Grey areas represent potential temporal bins for transduced signals. The first two bins represent the noise level. Blue data points are the mean photon counts with the error bars representing Poissonian uncertainty over 100 cycles, in the respective bin. \textbf{c,} Mean photon counts in bins for different numbers of temporal modes. \textbf{d(i),} Schematic of Fourier spectra of two pump pulses symmetrically detuned from the offset centre of the RAP spectrum. \textbf{d(ii),} Sweep of the pump detuning with MW frequency fixed at the spin resonance. The inset shows the corresponding temporal histogram. \textbf{e(i),} Schematic of two RAP windows placed around the centre of the optical inhomogeneous broadening. \textbf{e(ii),} 10 spectro-temporal modes stored in 4-memory cells defined by two pump frequencies ($\pm$) in two RAP windows ($\alpha /\beta$) each. Bit strings of 1's and 0's correspond to ON and OFF states of the MW in each mode while dash denotes OFF state for both pump and MW, and correspond to noise. Spectral coordinates are mentioned as (RAP $\alpha/\beta$, Pump $+/-$).} \label{fig:multiplexing}
\end{figure*}

Highly multimode QMs significantly enhance the entanglement generation rate in a repeater protocol~\cite{simon-2007}. Here, we demonstrate spectrally and temporally multiplexed M2O transduction, capitalizing on the large optical and spin inhomogeneous broadening in Yb:YSO as illustrated in Fig.\,\ref{fig:multiplexing}\textbf{a}.\par

\subsubsection{Temporal Multiplexing}
\label{sec:temporal_multiplex}
Here we transduce ten modes with all pump pulses at the same frequency and MW pulses at the spin resonance, rephased by a pair of \textcolor{black}{spectrally identical} single-tone RAP pulses in a First-In-First-Out (FIFO) sequence. In Fig.\,\ref{fig:multiplexing}\textbf{b}, the photon counts in each grey bin spaced by $5.4\,\mu$s, correspond to the temporal modes, where the first two labelled bins represent the noise counts. The transduction efficiency varies over the modes with the highest for the first mode, decreasing over the subsequent modes, and then increasing again for the final few modes. This behaviour is explained by first pump pulse interacting with the largest ground-state population, which enables a stronger population transfer to the $\ket{2_e}$ state. For the subsequent pulses, the available population in $\ket{4_g}$ decreases, leading to reduced efficiency. However, given the excited-state lifetime of approximately 1.3\,ms~\cite{welinski-2016} and a total multiplexing duration of only $100\,\mu\mathrm{s}$, some atoms repopulate the $\ket{4_g}$ state before the arrival of the last pump pulses, resulting in a partial recovery of transduction efficiency towards the end.\par

In order to better understand the effect of temporal multiplexing on efficiency, the experiment is performed with varying number of temporal modes as shown in Fig.~\ref{fig:multiplexing}\textbf{c}. A similar trend of decreasing efficiency over the modes is observed. The transduction efficiency of the first mode also decreases with increasing number of temporal modes. This is because of congestion of modes going to the same spectral window in the optical inhomogeneous broadening and can be overcome by optical spectral multiplexing.\par

\subsubsection{Opto-Spectral Multiplexing within Single RAP Window}
\label{sec:pump_spectral_multiplexing}
We utilize the broad chirp bandwidth of the RAP pulses ($\Delta_{\rm{ch}}=1.5$~MHz) to rephase two spectrally distinct pump modes with the same individual Fourier bandwidth of 250\,kHz ($\tau_p =4\,\mu\mathrm{s}$), as illustrated in Fig.~\ref{fig:multiplexing}\textbf{d(i)}. The pump frequency for the early and the late temporal modes are blue- $(+)$ and red-shifted $(-)$ respectively, by $\Delta\omega_P <\Delta_{\rm{ch}}/2$ from the offset centre of the RAP window such that $\omega_P^\pm = \omega_R + \omega_M \pm \Delta \omega_P$ define two spectral cells. The transduced counts for the early and the late modes are plotted with varying $\Delta\omega_P$ in Fig.\,\ref{fig:multiplexing}\textbf{d(ii)}. 
From temporal multiplexing, we know that the first temporal pulse supports the highest efficiency. 
This efficiency decreases with $\Delta\omega_P$ due to reducing efficiency of the rephasing by the RAP pulses. 
However, the late transduced counts with increasing $\Delta \omega_p$ initially rise, due to the decreasing overlap of the spectrum of the late pump pulse with that of the early pump pulse, before dropping again due to the inefficient RAP rephasing.
\textcolor{black}{We find that $\Delta\omega_P =200$\,kHz is optimal for balancing the efficiencies without sacrificing overall efficiency.}
Therefore, we can transduce pulses in any temporal order with the spectral cells symmetrically separated by 400\,kHz within a single RAP window.\par

\subsubsection{Opto-Spectral Multiplexing with multiple RAP Windows}
It is also possible to exploit the broad optical inhomogeneous broadening of Yb:YSO ($\sim$ 550\,MHz from SI~Fig.~S2~\cite{gautam2025spp}) to enable multi-window spectral multiplexing. Fig.~\ref{fig:multiplexing}\textbf{e(i)} illustrates two RAP windows ($\alpha$ and $\beta$) centred within the optical inhomogeneous profile, and can be extended to accommodate additional windows. Each RAP window can independently accommodate at least two spectral cells corresponding to different pump frequencies. Furthermore, within each pump frequency mode, multiple MW frequencies lying within the spin inhomogeneous broadening can be addressed (SI~Sec.~V(A)~\cite{gautam2025spp}), thereby enhancing the overall multiplexing capacity. An additional advantage of using multiple RAP windows is the ability to selectively retrieve stored modes from any chosen window at different times. While this approach is conceptually similar to the spectral multiplexing discussed in~\cite{kamel-2025}, here all the modes are retrieved together for simplicity.\par

For experimental demonstration, we define two RAP windows separated by 4\,MHz, each containing two cells, established by applying two pump frequencies detuned by $\Delta\omega_P=200\,\rm{kHz}$ from the offset centre of their respective RAP windows. 
In total, this configuration yields four distinct spectral cells, each capable of storing multiple MW spectral modes. The cells are named after their corresponding RAP and pump frequencies e.g. $\alpha -$ represents RAP$\alpha$ and red detuned pump pulse.
Fig.~\ref{fig:multiplexing}\textbf{e(ii)} presents the experimental results for ten spectro-temporal modes, where the bit-strings in the legend represent the ON (1) or OFF (0) state of MW pulse in the respective temporal mode. The pump pulses are ON for all the modes. In any particular temporal bin, the transduced counts are same if the MW is ON and it falls to the noise level if the MW is OFF. Therefore, a discriminator level shown by the horizontal dashed line can be used to determine the state of MW for that particular temporal bin. The transduction efficiency of the first four modes is higher than the next group of four modes due to the temporal-mode congestion within the same spectral cell. These observations highlight that distributing temporal modes across multiple RAP windows and pump frequency cells supports almost uniform efficiency across modes, demonstrating the effectiveness of spectro-temporal multiplexing. Note that the reduced counts in the even modes are not consistent in all the experiment trials, but an effect of the spatial inefficiency of the AOM at the shifted frequency (another trial is presented in SI~Fig.~S13~\cite{gautam2025spp}).\par   

%%%%%%%%%%%%%%%%%%%%%%%%%%%%%%%%%%%%%%%%%%%%%%%%%%%%%%%%%%%%%%%%%%%%%%%%%%%%%%%%%%%%%%%%%
\section{Discussion and Outlook}
\label{subsubsec:Discussion_and_outlook}
In this work, we present a memory-assisted M2O transduction protocol along with its implementation in Yb:YSO placed in a 3D LGR. The internal transduction efficiency, at $460\,\mu \textrm{s}$ storage time and $641\,\rm{kHz}$ bandwidth, is $4\times 10^{-8}$ with $\sim 0.4$ photons as noise floor.
We show coherent information transfer from MW to optical modes,
facilitating phase-encoding schemes such as QPSK. We further show spectro-temporal multiplexing capability, useful for time-bin qubit applications. Such a memory-assisted transducer with multiplexing is an important enabler for fast and efficient entanglement generation between SQCs.\par

To be clear of noise from FID decay from the RAP pulses and undesired echoes, extended storage times are required, impacting the M2O transduction efficiency. Introducing a small angle between the RAP and the pump pulses would suppress the noise and combined with a higher pump Rabi-frequency, using higher optical power, would allow higher efficiency, e.g., a $3.7$ times increase at $150\,\mu\mathrm{s}$ storage time. 
In the optical domain, a broader chirp range of the RAP pulses (only limited by the optical inhomogeneous broadening), would allow higher spectral multiplexing capacity.
Naturally, this will necessitate higher laser power to drive multiple RAP transitions simultaneously.\par

The low efficiency of our system is due to several factors, including low concentration of spins in the mode volume, small Rabi frequency at the pump transition, single-pass optical configuration, small optical-MW mode overlap, and low LGR quality factor. These factors result in weak MW and atomic cooperativities and similar efficiencies has been observed for Er:YSO without an optical cavity ($\sim10^{-12}$)~\cite{fernandez-gonzalvo-2015} and rises to $10^{-5}$ with an optical cavity~\cite{fernandez-gonzalvo-2019}. A planar superconducting MW resonator together with an optical cavity can provide a small mode volume with increased mode overlap, thereby enhancing the optical Rabi frequency, as well as large MW and atomic cooperativities, as demonstrated with Yb:YVO~\cite{xie-2025}. However, the short microwave/spin and optical memory coherence times ($T_{m,s} = 910\,$ns, $T_{m,o} = 140\,$ns)~\cite{xie-2025} makes Yb:YVO less suited for memory-assisted transduction. \textcolor{black}{With Yb:YSO in 3D-LGR and optical cavity under optimized conditions, the CW transduction efficiency can reach up to $10^{-4}$, and the memory efficiency to unity.} \par

\textcolor{black}{The memory-assisted transducer benefits from the long optical coherence time and high doping concentration (improves cooperativity) in REI-crystals. However, higher doping concentration decreases the optical coherence time~\cite{chiossi-2024}. Therefore, based on specific applications, the coherence time can be traded-off with high doping concentration to enable either long storage or high efficiency memory-assisted transduction. Higher doping also result in broader inhomogeneous linewidths, and can improve the MW bandwidth of the transducer~\cite{xie-2025, fernandez-gonzalvo-2019}.}

The protocol presented in this paper can be implemented in other systems, with the necessary coherence time and bandwidth, allowing for flexibility in matching wavelengths to other components over a link between remote SQCs.
Furthermore, our system can provide a valuable microwave-qubit memory interface for quantum computers. Even for purely optical QM realizations, the enhanced MW coupling of our system can be used to implement dynamical-decoupling sequences to increase the storage time.\par

%%%%%%%%%%%%%%%%%%%%%%%%%%%%%%%%%%%%%%%%%%%%%%%%%%%%%%%%%%%%%%%%%%%%%%%%%%%%%%%%%%%%%%
\section{Methods}  
\label{sec:Methods}

\subsection{Experimental Setup}
The experimental setup is presented in SI~Fig.~S1~\cite{gautam2025spp}. A Toptica DL Pro laser is locked to an external reference cavity (Stable Laser System) at 1150\,MHz blue-detuned from $|4_g\rangle \leftrightarrow |1_e\rangle$ transition. The laser beam passes through two acousto-optic modulators (AOMs). AOM1 (Model: AA Opto-Electronic: MT200-B100A0,5-1064) is used to create optical pulses and fine frequency shifts (0-5\,MHz). AOM2 (Model: Brimrose: TEM-85-10) suppresses the first-order leak light from AOM1, helpful for single photon detection with minimal background noise. A phase modulator (PM) (Model: iXblue: NIR-MPX950) is used to coarse shift ($\sim \rm{GHz}$) the frequencies for optical pumping, pump, and RAP pulses. The limited shift efficiency of the PM (70\%) creates undesired noise (2PE with leak light) and limits the pump pulse power. Optical powers mentioned in the text account for this inefficiency. Two polarization controllers (half-wave plates) are placed before and after the PM for efficient frequency shift and for aligning the polarization along the $\vec{D}_2$-crystal axis to achieve maximum absorption~\cite{tiranov-2018}, respectively. Optical coupling efficiency through the crystal, including fibre and splice losses, is about $10 \%$. The output fibre is coupled to an avalanche photo detector (APD - Model: PerkinElmer SPCM-AQRH-14-FC) via a free-space gating AOM (AOM3 - Model: AA Opto-Electronic: MT200-B100A0,5-1064) with $40 \%$ shift efficiency. The APD detection efficiency is $12\%$ at 980 nm.\par

MW pulses are generated from an Arbitrary Waveform Generator (AWG - Tektronix - AWG70002A) and gated via an electronic switch (Model: ZASWA-2-50DRA+) before entering the dilution fridge. The overall loss of the MW line from the AWG to the input port of the cavity is -15.58 dB, including losses in the switch, isolators, and the transmission lines.
\par

\subsection{Experimental Sequence}
Due to the intense pulses involved in the transduction sequence, the population in the three-level system changes. To provide a consistent initial population, an optical pumping sequence is applied. The population is initialized to $|4_g\rangle$ level by applying a set of chirped pulses around $|3_g\rangle \leftrightarrow |2_e\rangle$ and $|2_g\rangle \leftrightarrow |1_e\rangle$ transitions. After optical pumping, the population is allowed to relax to the ground state for $1\,\mathrm{s}$. This also allows the local heat to dissipate. Optical depth of about 1.4 is reached after population initialization as shown in SI~Fig.~S2~\cite{gautam2025spp}. Then the transduction protocol is applied as shown in Fig.\,\ref{fig:protocol} and outlined in Sec.~\ref{sec:protocol}. The gating AOM is shut throughout the experiment except (i) between the pump pulse and RAP1, (ii) between RAP1 and RAP2, and (iii) between RAP2 and RAP echo. After the transduction protocol ends, we wait for another 1\,s before restarting the sequence to allow the excited state population to decay back to the ground levels.\par

\subsection{Pulse Shaping}
The optical pulses are temporally shaped with AOM1, and its coarse ($\sim$GHz) and fine ($<5$\,MHz) frequencies are adjusted with the PM and AOM1, respectively. AOM2 is turned ON with a delay of 400\,ns to account for signal propagation delay. The RAP pulses employed in this work have sinc temporal profile with linearly chirped frequency, similar to that used in the QM paper~\cite{kamel-2025}. 
\begin{equation}
    P(t)= A(t) \sin (\omega_R t + \varphi (t)),
    \label{eq:0}
\end{equation}
with the phase and amplitude terms given by
\begin{align}
    A(t)&= A_0 \rm{sinc}\left(\frac{2(t-t_0)}{\tau_R} \right), \\  
    \varphi (t)&= \frac{\Delta_{\rm{ch}}}{\tau_R} (t-t_0)^2 \ ,
    \label{eq:1}
\end{align}
where, $A_0$ is an amplitude factor, $\tau_R$ is the duration, $\Delta_{\rm{ch}}$ is the chirp bandwidth and $t_0$ is the center of the RAP pulse.

The MW pulse frequency and phase are defined from the AWG using the function $A(t) \sin(2\pi \nu_M t + \phi)$, where $A(t)=\rm{sinc}(2t/\tau_M)$ is the temporal envelope of the pulse, $\nu_M$ is the frequency of the MW, $\phi$ is the phase and $\tau_M$ is the duration of the MW pulse. We employed sinc temporal profiles ($\rm{sinc}(2t/\tau_{P,M})$) for both pump and MW pulses throughout, except for spin inhomogeneous broadening characterization in SI~Fig.~S7\textbf{(c)}~\cite{gautam2025spp}, where both the pulses are square in shape. For interference experiments, we vary the phase of the late MW pulse relative to the early one. For Homodyne measurements, the LO pulses are also sinc-shaped with the minimum possible amplitude from the AWG to keep the photon counts per pulse to the lowest. Lower counts in the LO pulses ensure high visibility of the interference patterns. \par

For multiplexing experiments, the MW pulse power was high ($\sim 2\times10^{11}$ photons per pulse for temporal modes; $\sim 1\times10^{11}$ photons per pulse for opto-spectral modes) for a better SNR over all the temporal modes to counter the lower efficiency at the longer storage durations employed. Also, the temporal gap between each mode is set to resolve the individual peaks. For demonstration with multiple RAP windows, we used AOM1 to shift the RAP central frequency by $\pm 2\,\rm{MHz}$ on either side of the AOM1 drive frequency. The $4\,\rm{MHz}$ frequency separation is enough to avoid cross-talk between the cells in different RAP windows. At smaller frequency separations, the transduction efficiency of the modes is lower than maximum. If there is no separation, the RAP power will be just doubled, causing weak rephasing across all the modes due to larger-than $\pi$-pulse area for the RAP pulses.\par

\subsection{Choice of Transitions}

\textcolor{black}{Due to the weak branching ratio of 0.04~\cite{businger-2020} at the pump transition $\ket{4_g} \leftrightarrow \ket{2_e}$, we could achieve only a $0.009\,\pi$-pulse, instead of a $\pi/2$-pulse required for maximum coherence (SI~Sec.~II(A)~\cite{gautam2025spp}), with a sinc-shaped pump pulse with Rabi frequency $\Omega_P = 2\pi\times 7\,\mathrm{kHz}$ at 2.5\,mW optical power. In contrast, the significantly stronger branching ratio of $0.72$~\cite{businger-2020} at $\ket{4_g}~\leftrightarrow~\ket{1_e}$ transition enables efficient rephasing with the RAP pulses at $850\,\mu\mathrm{W}$ peak optical power ($\Omega_R = 2\pi\times 75\,\mathrm{kHz}$) (SI~Fig.~S7\textbf{a}~\cite{gautam2025spp}). The corresponding adiabaticity condition $\Omega_R^2/R_{\rm{ch}}=1.41 >1$ is satisfied. The MW frequency sweep characterizes the spin resonance to be $2623.35\,\mathrm{MHz}$ with a bandwidth of $\Gamma_s = 642\,\mathrm{kHz}$ (SI~Fig.~S7\textbf{c}~\cite{gautam2025spp}), matching well to the Fourier bandwidth of the MW pulses.}\par

\subsection{Efficiency Budget} \label{subsec:efficiency_budget}
\textcolor{black}{The memory-assisted transduction efficiency ($4\times 10^{-8}$) depends on the memory and continuous-wave (CW) transduction efficiencies. The optical memory efficiency measured using the same pulse parameters is approximately 7\%. The CW transduction efficiency is measured to be $\sim 1\times10^{-8}$ under comparable pump powers, albeit with slightly different field configurations. In the CW measurement, the optical pump is applied on the $|4_g\rangle \leftrightarrow |1_e\rangle$ transition, whereas in the memory-assisted protocol the pump addresses the $|4_g\rangle \leftrightarrow |2_e\rangle$ transition. As a consequence, in the CW case the generated optical field at the $|4_g\rangle \leftrightarrow |2_e\rangle$ transition spectrally overlaps with the inhomogeneously broadened $|2_g\rangle \leftrightarrow |1_e\rangle$ transition, leading to partial reabsorption and attenuation of the transduced optical signal. In addition, the CW transduction experiment employs longer and stronger microwave and optical fields, resulting in lower efficiency due to saturation of the spin transition~\cite{rochman-2023}.}
%---------------------------------------------------------------------------------------%
\begin{acknowledgments}
%The authors would like to thank ... for valuable discussions on this research.
This work was supported by the Government of Alberta Major Innovation Fund Project on Quantum Technologies, Alberta Innovates Advance Grant, Mitacs through the Mitacs Accelerate program, the Canadian Foundation for Innovation Infrastructure Fund (CFI-IF), and the Natural Sciences and Engineering Research Council of Canada (NSERC) through the Alliance Quantum Consortia Grants CanQuest and ARAQNE.
\end{acknowledgments}

%
%%
%%%
%%%%
%%%%%
%%%%%%
%%%%%%%
%%%%%%%%
%%%%%%%
%%%%%%
%%%%%
%%%%
%%%
%%
%

%\cleardoublepage
\onecolumngrid
\appendix
\newpage

\section{Experimental Setup}\label{app:setup}
\begin{figure*}[htbp]
    \centering  \includegraphics[width=1\linewidth]{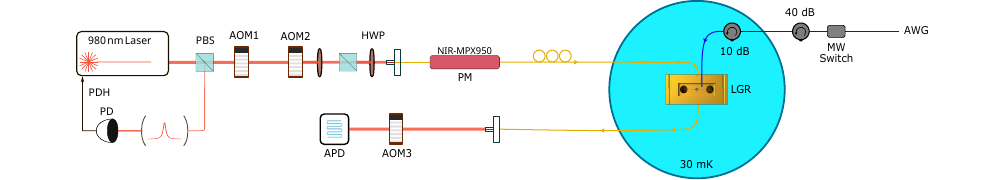}
    \caption{\textbf{Experimental Setup.} PBS - Polarizing Beam Splitter; PD - Photo diode; PDH - Pound-Drever-Hall; AOM - Acousto-optic modulator; HWP - Half-wave plate; PM - Phase modulator; APD - Avalanche photodiode; LGR - Loop-gap resonator; 10/40 dB isolators; AWG - Arbitrary waveform generator. Red lines represent free-space beams, yellow lines are optical fibres, and blue lines are MW lines.}  
    \label{fig:Setup}
\end{figure*}

\subsection{Optical Depth}
\begin{figure*}[h!]
    \centering  \includegraphics[width=1\linewidth]{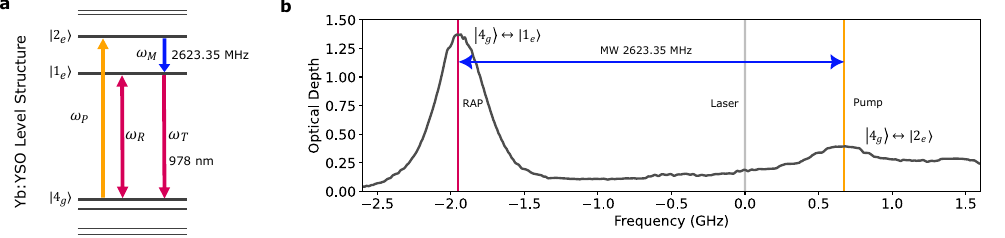}
    \caption{\textbf{Optical Depth Spectrum.} \textbf{a,} Energy level structure of Yb:YSO with optical and MW transitions highlighted with different colours. \textbf{b,} Spectrum of $\ket{4_g} \leftrightarrow \ket{1_e}$ and $\ket{4_g} \leftrightarrow \ket{2_e}$ transitions after a sequence of optical pumping to initialize maximum population in $\ket{4_g}$ level. Vertical lines mark the optical frequencies, with grey, yellow and red lines representing laser frequency just before the PM shift, pump frequency and RAP frequency respectively. The RAP frequency aligns with the $\ket{4_g} \leftrightarrow \ket{1_e}$ transition while the pump frequency aligns with the $\ket{4_g} \leftrightarrow \ket{2_e}$ transition. The height of the peaks are related to their respective branching ratios. Zeroth order of the PM interacts with a very small OD and causes the noise discussed in the main text.}  \label{fig:OD_plot}
\end{figure*}

%\newpage
\twocolumngrid

\section{Theoretical Model}
\label{SI:Theory}

\begin{figure}[htbp]
    \centering  \includegraphics[width=0.65\columnwidth]{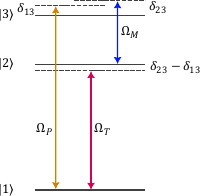}
    \caption{\textbf{Three-level System.} Pump field drives $\ket{1}\leftrightarrow\ket{3}$ transition with detuning $\delta_{13}$, MW field drives the excited state $\ket{2}\leftrightarrow\ket{3}$ transition with detuning $\delta_{23}$, RAP and signal fields are driven and emitted at $\ket{1}\leftrightarrow\ket{2}$ transition with detuning $\delta_{23} - \delta_{13}$ to maintain the energy conservation.}  
    \label{fig:Level_diagram}
\end{figure}

QM using RAP pulses is described by an ensemble of two-level systems~\cite{kamel-2025}. Here, we extend this idea to an ensemble of three-level systems for memory-assisted transduction. In Fig.\,\ref{fig:Level_diagram}, a three-level system is shown with one ground and two excited states. The V-arms of the system $\ket{1}\leftrightarrow\ket{3}$ and $\ket{1}\leftrightarrow\ket{2}$ are driven by optical fields, while the excited state spin transition is driven by a MW field. The system Hamiltonian in rotating frame is given by the following set of equations:

\begin{align}
\begin{split}
\hat H_0 &= (-\delta_{13}+\delta_{23}) \hat\sigma_{22}-\delta_{13} \hat\sigma_{33},\\
\hat H_{I, P} (t) &= \frac{\Omega_{P}(t)}{2} (\hat\sigma_{13} + \hat\sigma_{31})\\
\hat H_{I, M} (t) &= \frac{\Omega_{M} (t)}{2} (\hat\sigma_{23} + \hat\sigma_{32})\\
\hat H_{I, R} (t) &= \frac{\Omega_{R} (t)}{2} \left( e^{-i \phi_R(t)} \hat\sigma_{12} + e^{i\phi_R(t)} \hat\sigma_{21} \right),\\
\hat H &= \hat H_0 + \hat H_{I, P} + \hat H_{I, M} + \hat H_{I, R}.
\end{split}
\label{eq:Hamiltonian}
\end{align}

where, $\hat H_{I, P}$, $\hat H_{I, M}$, and $\hat H_{I, R}$ are interaction terms corresponding to the pump, MW and RAP pulses respectively. Detunings $\delta_{13}$ and $\delta_{23}$ are shown in the level diagram (Fig.~\ref{fig:Level_diagram}). $\hat\sigma_{ij}= \vert i\rangle \langle j\vert$ for $i, j \in \{1,2,3\}$ define the transition operators. $\Omega_P$, $\Omega_{M}$ and $\Omega_R$ are the Rabi frequencies corresponding to the pump, MW and RAP pulses respectively. Phase $\phi_R(t)$ is due to the RAP frequency chirp. Each interaction component is active only when those pulses are applied e.g. $\hat H_{I, P}$ is active only during the pump pulse.\par

\begin{figure*}[htbp]
    \centering  \includegraphics[width=0.9\linewidth]{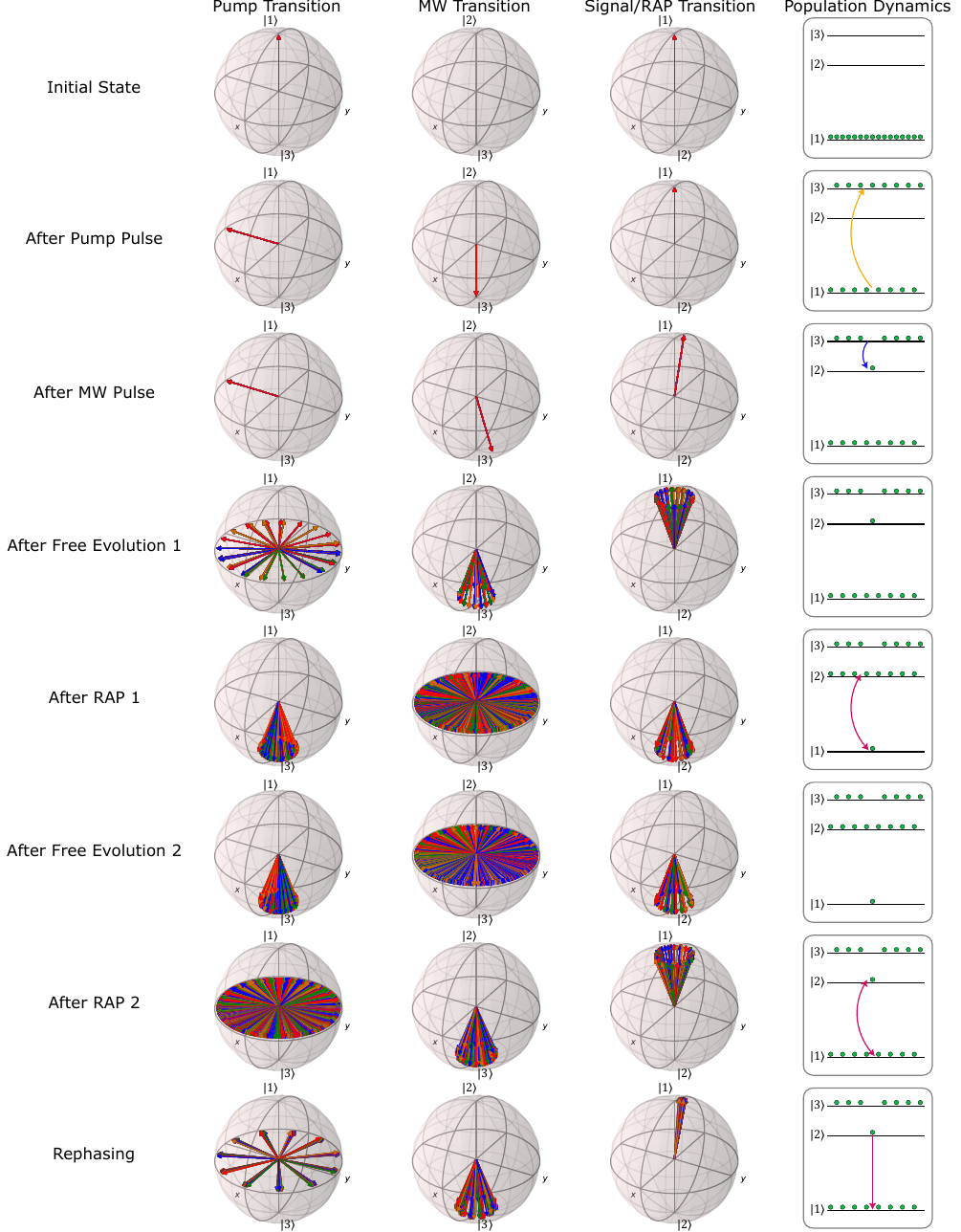}
    \caption{\textbf{Evolution of state vectors.} Pump, MW and RAP transitions are three two-level sub-systems. State vectors defined by detunings in the inhomogeneous broadening are shown at different stages of the protocol mentioned in the left column. Color of the vectors have no physical significance. Vectors during the dephasing appears to be bunched but is just an aliasing effect. Population dynamics after each stage is shown on the right-most column.}  \label{fig:Bloch_spheres}
\end{figure*}

\begin{figure*}[t]
    \centering  \includegraphics[width=1\linewidth]{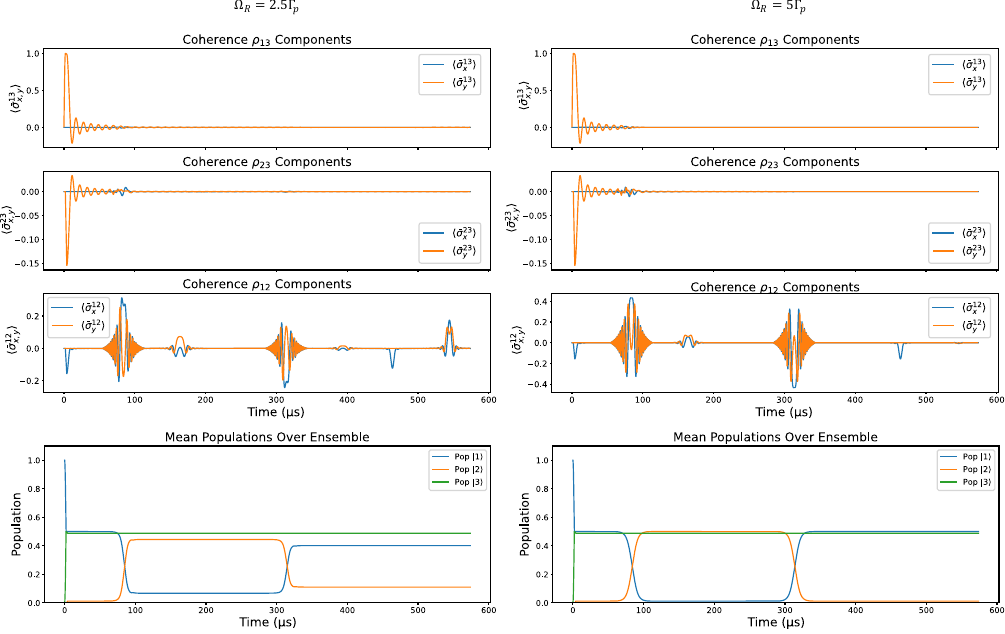}
    \caption{\textbf{Simulation of Coherence and Population Transfer.} Evolution of coherences (top panels) and population (bottom panel) on various transitions for a given Rabi frequency of the RAP pulses, $\Omega_R$. Left (right) column is with $\Omega_R =2.5 \Gamma_P (5 \Gamma_P)$, where $\Gamma_P$ is the FWHM of the pump transition (30\,kHz) used in the simulations. $\langle \bar\sigma^{13}\rangle$ - Pump transition; $\langle \bar\sigma^{23}\rangle$ - MW/spin transition; $\langle \bar\sigma^{12}\rangle$ - RAP/signal transition. RAP pulses rephase the coherence at a later time with unit efficiency in a lossless model. Undesired echos are better suppressed at higher $\Omega_R$ because of better adiabaticity conditions. The last peak in $\langle \bar\sigma^{12}\rangle$ the left column plot is the RAP echo and is also suppressed when $\Omega_R$ is higher (barely seen in the right column plot).}
    \label{fig:coherence_pop}
\end{figure*}

The time dependent Hamiltonian (eqn.~\ref{eq:Hamiltonian}) describes the evolution of individual three-level systems in the ensemble. To describe the interaction in the ensemble picture, inhomogeneous broadenings are considered in both the optical and spin domains. We solve the time dependent Hamiltonian using QuTiP in a $m\times m$ grid defined by detunings $\delta_{13}$ and $\delta_{23}$. We consider weight factors based on the detunings in a Gaussian (for optical inhomogeneous broadening) or Lorentzian (for spin inhomogeneous broadening) distribution. The simulation's computational space complexity grows with grid size $m\times m$. Therefore we restrict $m$ to 150 for our simulation results. For sufficient resolution, we constraint the inhomogeneous broadenings to narrow bandwidths i.e. 30 kHz for each broadening.\par

All the $m\times m$ (=22500) state vectors corresponding to each detuning is plotted on Bloch spheres of the two-level sub-systems in Fig.~\ref{fig:Bloch_spheres}. Each vector is the final state for that detuning, after the application of pulses or free evolution mentioned on the left column (see main text Fig.1\textbf{(c)}~\cite{gautam2025memory_trans_main} for pulse sequence). The corresponding population transfer dynamics is shown on the right column. In each Bloch sphere, the vectors are normalized to avoid showing vector shortening or stretching due to population changes in that transition. Mathematically, if the state of the three level system is $\ket{\psi_{123}}=c_1\ket{1} + c_2\ket{2} + c_3\ket{3}$, then the normalized state vector in a sub-system is $\ket{\psi_{ij}}=(c_i\ket{i} + c_j\ket{j})/\sqrt{c_i^2+c_j^2}$, where $i,j\in \{1,2,3\}$ and $i\ne j$.\par

In Fig.\,\ref{fig:Bloch_spheres}, the north pole of the Bloch spheres is the lower level in their respective two-level sub-systems. With the initial population in the ground level, the initial state vectors point upwards in both the pump and RAP transitions. Since, there is no population in the excited state yet, the Bloch sphere for MW transition has no vectors (Fig.\,\ref{fig:Bloch_spheres} Initial State, MW Transition). For simulations, we consider a $\pi/2$ pump pulse, which rotates the vectors in the pump transition by $\pi/2$ about the x-axis. This also means that some population is transferred to the excited state which reflects in Bloch vectors appearing in the Bloch sphere for MW transition (Fig.\,\ref{fig:Bloch_spheres} After Pump Pulse, MW Transition). 
Note that the Bloch vectors in the RAP transition do not move because the population was transferred only to the second excited state ($\ket{3}$). 
After a weak MW pulse ($\pi/10$), the vectors in the MW transition Bloch sphere rotates by $\pi/10$ about the x-axis. We choose a $\pi/10$ rotation to highlight its effect on the Bloch sphere. However, a single photon excitation will generate coherence in a similar fashion. Due to the application of the MW pulse, a small rotation is also seen in the RAP transition Bloch vectors. This weak coherence in the RAP transition is similar to the coherence excited by an optical probe pulse in the Chirped Pulse Phase Imprinting (CPPI) QM protocol~\cite{kamel-2025}. The rest of the pulse sequence is the same as for CPPI protocol on the RAP transition while the Bloch vectors on the pump and the MW transitions evolve freely.\par

During the free evolution 2 i.e. between RAP1 and RAP2, the Bloch vectors in the RAP transition dephase closer to the excited state $\ket{2}$. After time $\tau_1$, these vectors weakly rephase giving rise to the undesired echo (see main text Figs.~1\textbf{(d,e)}~\cite{gautam2025memory_trans_main} and Fig.~\ref{fig:coherence_pop}). This also occurs at time $\tau_1$ after RAP2. Notice that only the vectors in the RAP transition rephases while the vectors in the other transitions keep dephasing.\par

In order to clearly observe rephasing, we plot the mean value of collective coherences in Fig.~\ref{fig:coherence_pop} defined by:

\begin{align}
    \langle \overline\sigma_{x,y}^{i,j} \rangle=\frac{1}{N}\sum_{k=1}^N \langle \hat\sigma_{x,y}^{i,j} \rangle^{(k)}
    \label{eq:collective_coh}
\end{align}

where, $i,j\in \{1,2,3 \}$ represent the energy levels, $k$ is the index of the Bloch vectors over which the normalization is performed and N ($=m^2$) is the number of Bloch vectors used in the simulation. $x$ and $y$ denote the component of coherence along the $x$ and $y$-axis of the Bloch sphere. On the Bloch sphere, equation~\ref{eq:collective_coh} represents the mean value of projections of all the Bloch vectors along the $x$ and $y$-axis.\par

The mean values of the coherences and populations over the detuning grid are plotted over time in Fig.~\ref{fig:coherence_pop}. The pump pulse creates an average coherence in the pump transition. The MW pulse then creates coherence which also result in finite coherence in the RAP transition. The MW pulse weakly affects the populations in the levels. All these collective coherences dephase and average out to zero. To rephase this collective coherence a pair of RAP pulses are utilized. RAP1 pulse inverts the population while simultaneously imprinting phase on the Bloch vectors depending on their detunings. This additional imprinted phase is then cancelled out by RAP2 pulse which also brings the population closer to the ground state. RAP pulses with higher Rabi frequencies invert the population better (see right in Fig.~\ref{fig:coherence_pop}) and result in better silencing of undesired and RAP echos. This is due to improved adherence to the adiabaticity condition.\par

\subsection{Initial Coherence in a single-atom picture}
The state of the system starts with the ground state $\ket{1}$. It is transformed by the intense pump and weak MW pulse as:
\begin{equation}
    \ket{1} \xrightarrow{\theta_P} \alpha\ket{1}+ \beta\ket{3} \xrightarrow{\theta_M} \alpha\ket{1}+ \beta(\gamma\ket{2} + \ket{3})
\end{equation}
where, $\theta_P$ is the rotation angle due to the intense pump pulse and $\theta_M$ is the rotation angle due to the weak MW pulse. Therefore, $\gamma$ is a small complex coefficient compared to $\alpha$ and $\beta$. At the end of the pump-MW pulse sequence, the coherence $\langle \sigma^{12}\rangle$ is $\alpha^* \beta \gamma$. The maximum coherence in the RAP transition with a small $\gamma$ is achievable when $\alpha = \beta = 1/\sqrt{2}$ i.e. when the pump pulse is a $\pi/2$-pulse.

\subsection{Initial Coherence decay with Pump-MW delay}
\begin{figure}[h!]
    \centering  \includegraphics[width=0.95\columnwidth]{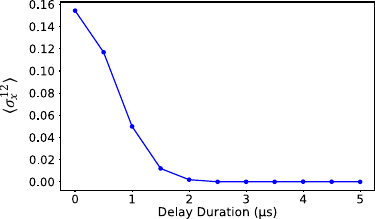}
    \caption{\textbf{Collective Coherence Decay with delay.} Simulation of coherence $\langle \sigma_{x}^{12}\rangle$ after pump-delay-MW pulse sequence plotted as a function of delay $\tau_0$.}  \label{fig:coherence_decay_SI}
\end{figure}

Any delay between the pump and the MW pulse results in weaker coherence transfer at the signal transition. We simulate this behaviour by considering spin (optical) inhomogeneous linewidth of 600\,kHz (1500 kHz) and $250\times 250$ atoms and vary the delay ($\tau_0$) between the pulses. The resulting plot (Fig.~\ref{fig:coherence_decay_SI}) reveals similar order of decay time $\sim 1\,\mu \mathrm{s}$, thereby confirming our hypothesis of spin dephasing induced decay profile.

%\clearpage

\section{Characterizations}
\subsection{RAP Power}
\begin{figure*}[htbp]
    \centering  \includegraphics[width=1\linewidth]{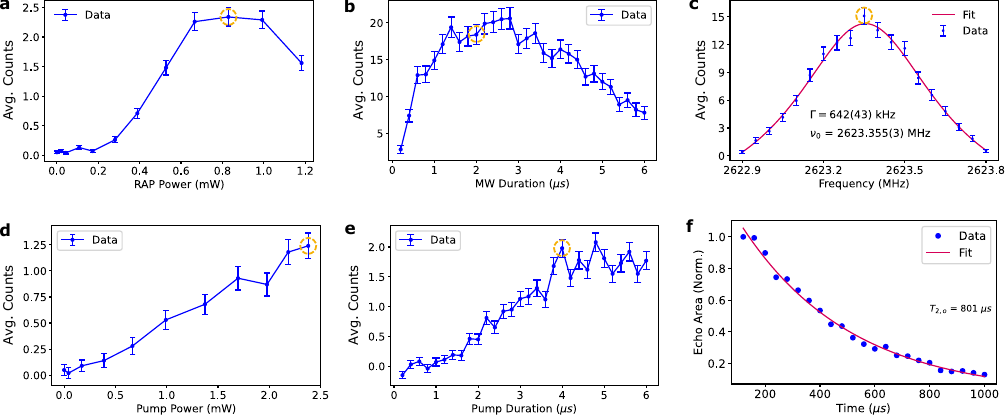}
    \caption{\textbf{Characterizations.} Each data point and its error-bar is the average and Poissonian uncertainty respectively, over 100-cycles (except in \textbf{f}). \textbf{a,} Optimization of RAP pulse peak power for maximum transduction efficiency. \textbf{b,} Sweep of MW pulse duration with square pulse shape showing optimal duration of $\sim 2\,\mu \rm{s}$. \textbf{c,} MW pulse frequency is swept across the spin transition. Data is fitted to a Lorentzian profile resulting in $642 \pm 43$ kHz linewidth and $2623.355 \pm 0.003$ MHz resonant frequency. \textbf{c,} Sweep of pump pulse power showing no signs of saturation. \textbf{e,} Sweep of pump pulse duration with sinc pulse shape showing optimal duration of $\sim 4\,\mu \rm{s}$ for a given RAP chirp bandwidth. \textbf{f,} Two-pulse echo data to measure the optical coherence time of $\ket{4_g} \leftrightarrow \ket{1_e}$ transition.}  \label{fig:Characterizations_SI}
\end{figure*}

The RAP pulse duration is fixed at $60\,\mu\textrm{s}$ with 1.5\,MHz chirp bandwidth. In Fig.\,\ref{fig:Characterizations_SI}\textbf{a}, the RAP power is swept and the corresponding transduced signal counts are plotted. The transduction efficiency is maximum at $850\,\mu$W peak optical power which corresponds to a peak Rabi frequency of $\Omega_R = 75\,\textrm{kHz}$. This is equivalent to a $\pi-$pulse characterization, since the RAP pulses invert the populations while imparting detuning dependent phase on the atoms. 
% \subsection{RAP Pulse Chirp Range and Pump Pulse Duration}
\subsection{Pump Pulse Power \& Duration}
After optimizing for RAP power, the pump pulse power is swept to see its effect on efficiency. Fig.\,\ref{fig:Characterizations_SI}\textbf{b} depicts that the efficiency keeps rising with increasing pump power without any sign of saturation.\par

The Rabi frequency at the signal transition with branching ratio 0.72 is measured to be $2\pi\times178\,\textrm{kHz}$ at 5\,mW optical power. Therefore, the Rabi frequency of the pump field with a branching ratio of 0.04 and $\sim 2.5$\,mW optical power is calculated by $2\pi\times178\,\textrm{kHz}\times \sqrt{2.5/5} \times 0.04/0.72 = 7\,\textrm{kHz}$.
Since the pump Rabi frequency is very small, its excitation bandwidth is limited by its Fourier bandwidth. Therefore, the pump Fourier bandwidth must be within the RAP Fourier window for optimal rephasing. The optimal pump pulse duration in Fig.\,\ref{fig:Characterizations_SI}\textbf{e} is $4\ \mu s \equiv 250 \, \textrm{kHz} = 1500/6 \,\textrm{kHz}$, where 1500\,kHz is the RAP chirp bandwidth. Square pulse shapes are used for pulse duration optimization with a square MW pulse duration set to $2\ \mu\,\textrm{s}$ at a constant amplitude. RAP parameters are also fixed for this experiment.\par

\subsection{MW Frequency \& Pulse Duration}
In Fig.\,\ref{fig:Characterizations_SI}\textbf{c}, we vary the MW frequency around the expected spin transition frequency. A Lorentzian fit to the data results in a linewidth of $\Gamma_s = 642 \pm 43$ kHz, centred at $2623.355 \pm 0.003$ MHz. Similar to the pump pulse case, the weak MW excitation is limited by its Fourier bandwidth. This bandwidth must fit in the spin inhomogeneous broadening for maximum absorption and therefore maximum transduction efficiency. In Fig.\,\ref{fig:Characterizations_SI}\textbf{d}, the transduced photon counts are plotted for varying MW pulse duration. The amplitude of the MW pulse is varied with its duration as $1/\sqrt{\tau_{M}}$, where $\tau_{M}$ is the MW pulse duration. This ensures constant number of input MW photons going into the LGR. The optimal MW pulse duration from the plot is about $2.2 \ \mu s \equiv 454 \ \text{kHz}$, which is well within the spin inhomogeneous broadening ($642\,\text{kHz}$) determined in Fig.\,\ref{fig:Characterizations_SI}\textbf{c}. Note that a square pulse shape is used for this characterization.

\subsection{Optical Coherence Time}
We apply 2-pulse echo (2PE) sequence to extract the optical coherence time of the RAP transition. The echo area is plotted with total delay time in Fig.\,\ref{fig:Characterizations_SI}\textbf{f}. An exponential decay function of the form $A\exp(-2t/T_{2,o})$ yields an optical coherence time of $T_{2,o} = 801\,\mu s$. The limited optical coherence time restricts the transduction efficiency at longer storage durations.

\subsection{Efficiency \& Calibration}
\begin{figure}[h!]
    \centering  \includegraphics[width=0.95\linewidth]{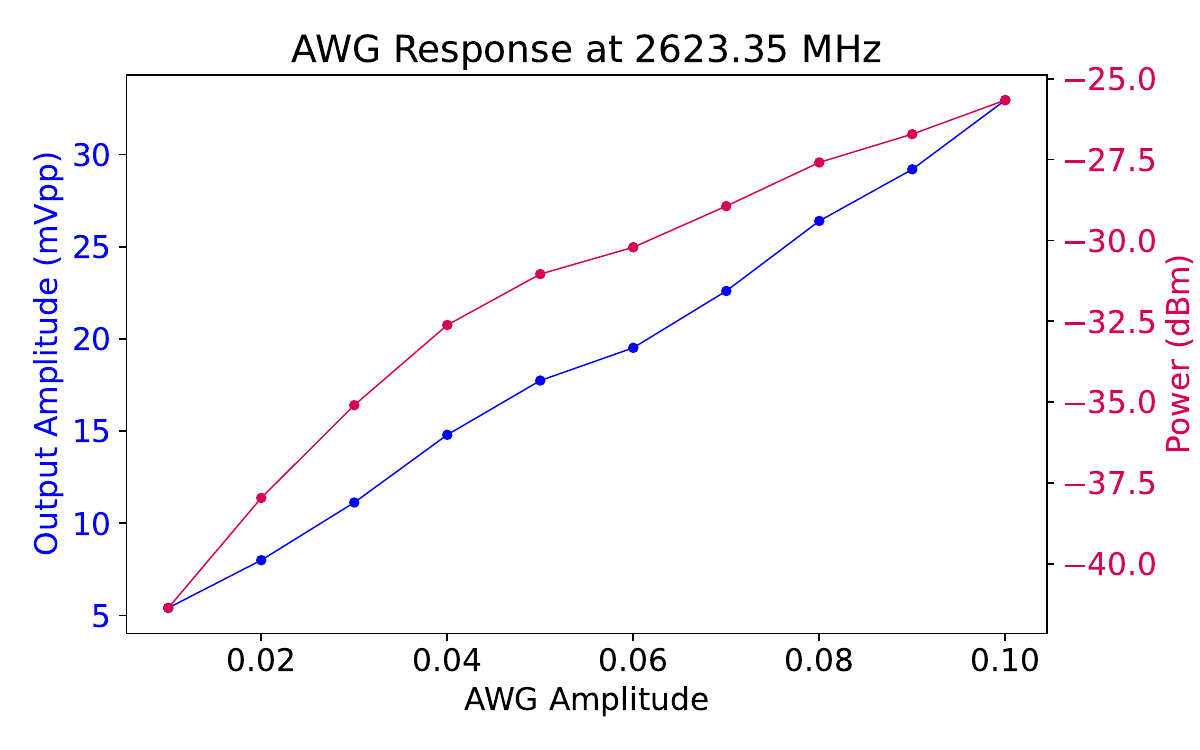}
    \caption{\textbf{AWG Power Output.} Peak-to-peak voltage of a sine wave at 2623.35\,MHz generated from the AWG at different AWG amplitude (ranges from 0 to 1) is measured on an oscilloscope near weak power outputs. The corresponding peak-to-peak voltage is converted to dBm on the right-y-axis. The power plotted here includes 2.5 dB loss in the MW cable used for connecting AWG to the oscilloscope. Typically 0.05 AWG amplitude is used to drive the MW pulses in the experiment.}  \label{fig:AWG_power}
\end{figure}

To characterize the photon conversion efficiency, we need to know the number of input MW photons at the input port of the LGR. Overall loss from AWG to the input port of the LGR is -15.58 dB. For most of the experiment, we send 23.66 mVpp corresponding to 0.05 amplitude from the AWG (Fig.\,\ref{fig:AWG_power}). This corresponds to -44.4 dBm power reaching the input port of the LGR. For MW frequency of 2623.35\,MHz, the number of photons contained in a square pulse of duration $2\ \mu s$ is $4.2 \times 10^{10}$. For a sinc pulse of $2\ \mu s$ duration, the number of photons is $2.5 \times 10^{10}$. The average number of transduced optical photons with a sinc profile pump pulse and the sinc MW pulse is 2 ($N_s - N_n$) (see Fig.\,\ref{fig:pulse_shapes}), resulting in total efficiency of $8\times 10^{-11}$. The total efficiency includes losses in the optical coupling ($10 \%$), gating AOM efficiency ($40 \%$), and APD detector efficiency ($12 \%$). This yields an internal photon conversion efficiency of $1.7\times 10^{-8}$.

\subsection{Pump and MW Pulse Shapes}
\begin{figure}[h!]
    \centering  \includegraphics[width=0.95\linewidth]{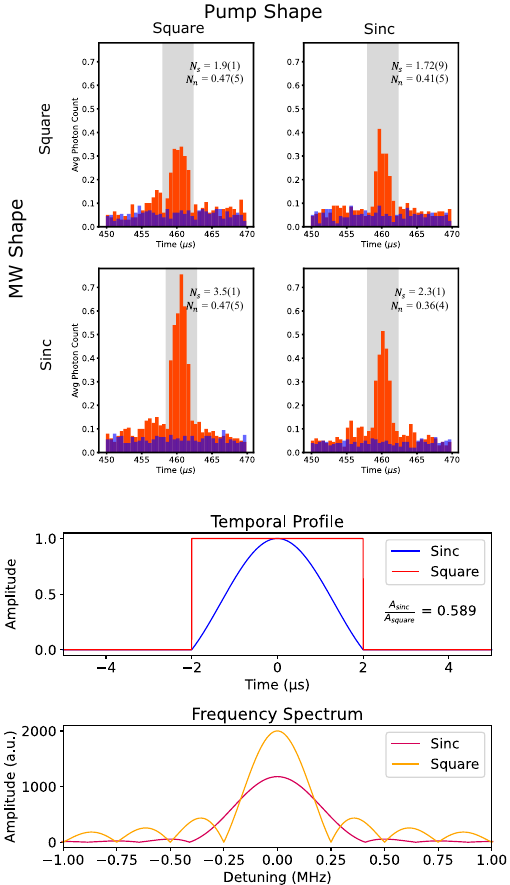}
    \caption{\textbf{Pump and MW Pulse shapes.} \textbf{Top,} Histogram of photon counts averaged over 200-cycles for different pump and MW pulse shapes. $N_s$ and $N_n$ are the transduced signal and noise photon counts respectively, in the grey detection bin. Input MW photons are the same in all the cases. \textbf{Bottom,} Temporal and spectral profiles of square and sinc pulse shapes. The area of a sinc pulse is $0.589$ times that of a square pulse with the same duration. However, the spectral sidebands in the square pulse creates noise, making sinc pulses preferable for low noise operations.}\label{fig:pulse_shapes}
\end{figure}

As discussed in sec.~II~E~\cite{gautam2025memory_trans_main} of the main text, the Fourier spectrum of the pump and MW pulses should match the RAP Fourier window and spin inhomogeneous broadening respectively. 
To figure out optimal shape, we send in two different temporal pulse shapes (square and sinc) and plot the histogram of the transduced photons in Fig.\,\ref{fig:pulse_shapes}. 
Area under a sinc pulse is $58.9\%$ of the area under a square pulse with the same pulse duration. This means that if a square pulse applies $\theta$ rotation, then a sinc pulse can apply only $0.589\,\theta$. Therefore, the coherence generated by the pump pulse will be higher for a square pulse compared to a sinc pulse of identical duration. However, the sidebands in a square pump pulse create more noise due to inefficient rephasing with the RAP pulses. Since the sidebands are minimal for a sinc pulse, the noise level is lower than that for square pulses. Therefore, we can choose between different pulse shapes and trade-off efficiency with noise level. Also, note that the bandwidth of the two pulse shapes are different and therefore optimal pulse duration will vary with the pulse shapes. Throughout the experiments, we have mostly used sinc pulse shapes for both pump and MW, whose pulse durations are slightly away from the optimal durations presented in Fig.\,\ref{fig:Characterizations_SI}\textbf{d}-\textbf{e}. In Fig.\,\ref{fig:pulse_shapes}, the amplitude of the MW pulses were adjusted to account for the pulse area difference.

\section{More on Interference}
\subsection{Bloch Vector - Ensemble Picture}

\begin{figure}[h]
    \centering
    \includegraphics[width=0.95\linewidth]{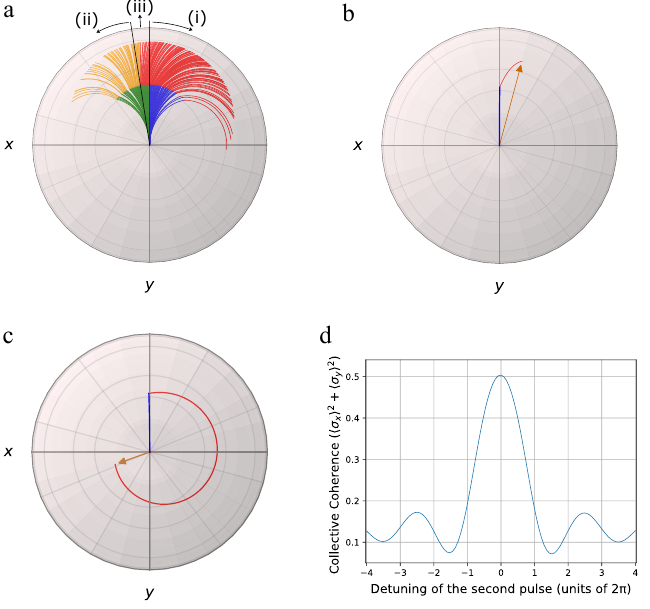}
    \caption{Ensemble Bloch vector rotation with a two-pulse sequence. \textbf{a,} Top view of the Bloch Sphere. Path taken by individual Bloch vectors with positive (green) and negative (blue) field detuning corresponding to the rotation due to the first pulse, while positive (yellow) and negative (red) field detuning corresponding to the second pulse. \textbf{b and c,} Ensemble average vector rotation with the second pulse detuned by $0.1\times 2\pi$ and $1.5\times 2\pi$. (d) Collective ensemble coherence as a function of detuning of the second pulse.}
    \label{fig:ensemble_bloch_vector}
\end{figure}

\textcolor{black}{Consider a single atom interacting with a monochromatic field. The atom-field interaction can be seen as a rotation of the Bloch vector, corresponding to the atomic state, around a precession axis or $\vec{p}$. The $\vec{p}$ is determined based on the phase and detuning of the field relative to the atom~\cite{foot-2005}. The 0-detuning ($\Delta\nu=0$) $\vec{p}$ lies on the $x$-$y$ plane making an angle $\phi$ (phase of the field) with the $x$-axis. For non-zero detuning the $\vec{p}$ has a $z$-component, thereby moving the $\vec{p}$ away from the $x$-$y$ plane. For instance, a 0-phase $\vec{p}$ with positive detuning has positive $x$ and $z$-components. Therefore, a rotation about this $\vec{p}$, takes the Bloch vector out of the $y$-$z$ plane. If we look at the projection of the rotated vector on $x$-$y$ plane from the positive $z$-axis, the net effect is a counter-clockwise rotation. Similarly, if a 0-phase field has negative detuning, the Bloch vector rotates clockwise w.r.t. the $z$-axis.}

\textcolor{black}{Now, consider an ensemble of atoms interacting with a monochromatic field resonant to the centre of the inhomogeneous broadening (see blue and green paths in Fig.~\ref{fig:ensemble_bloch_vector}\textbf{a}). The atoms negatively (positively) detuned from the centre will see a positive (negative) detuning field. Therefore, the field rotates the individual Bloch vectors corresponding to their relative detuning. In the ensemble picture, the effect of positive and negative detuning will cancel out, leaving the component along the 0-detuning component. This implies that we can treat the ensemble of atoms as a single atom when interacting with a resonant field.}

\textcolor{black}{However, the physics change when two such fields with finite detuning between them interact with the ensemble of atoms. To understand this, consider the schematic shown in Fig.\,4\textbf{b} of the Main text~\cite{gautam2025memory_trans_main}. Assume that the two fields are monochromatic at their respective centre frequencies. Now, we can define three zones in the atomic ensemble: (i) Atoms negatively detuned from both the fields, (ii) atoms positively detuned from both the fields, (iii) atoms positively detuned from one and negatively detuned from the other one. Case (i) and (ii) are easy to understand--the Bloch vectors in this region will rotate in the same clockwise or counter-clockwise direction w.r.t. z-axis (notice the paths after the two pulses in Fig.~\ref{fig:ensemble_bloch_vector}\,\textbf{a} going successively in the same direction either clockwise or counter-clockwise). The atoms in case (iii) will first rotate based on the negative detuning of the first field say counter-clockwise and then rotate clockwise due to the positive detuning of the second field (notice the paths after the two pulses in Fig.~\ref{fig:ensemble_bloch_vector}\,\textbf{a} going first in the counter-clockwise direction and then in the clockwise direction). Therefore, the ensemble average Bloch vector will follow a different path than if the pulses were not detuned. In Fig.~\ref{fig:ensemble_bloch_vector}\,\textbf{b} and \textbf{c}, the first pulse is kept resonant while the second pulse is detuned by $0.1\times 2\pi$ and $1.5\times 2\pi$, respectively. The length of the projected vector (or coherence) after the successive rotations determine the amplitude of transduced signal in Fig.\,4\textbf{c} and \textbf{d} of the Main text~\cite{gautam2025memory_trans_main}.}

\subsection{Interference Plots}
\begin{figure*}[h!]
    \centering  \includegraphics[width=0.95\linewidth]{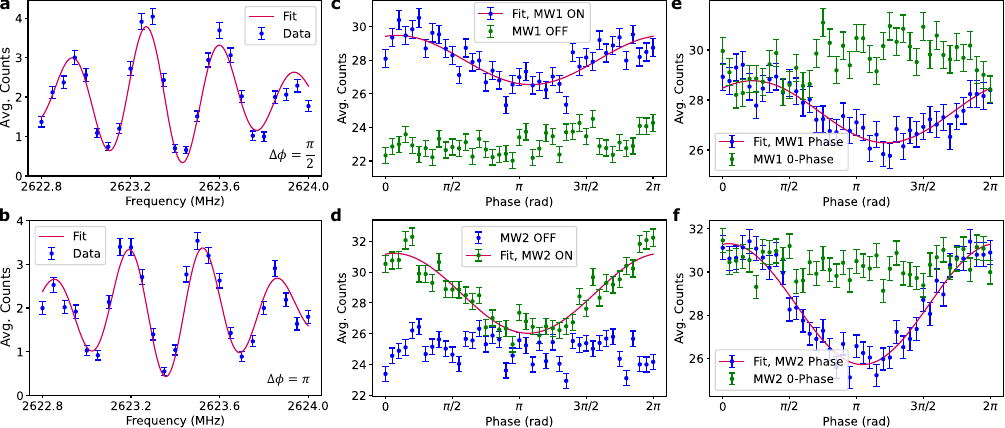}
    \caption{\textbf{Additional Data on Interference.} \textbf{a} and \textbf{b,} Interference plot with varying frequency of MW2 at $\Delta \phi = \pi/2 \ \text{and} \ \pi$, respectively, similar to main text Fig.~4\textbf{(d)}~\cite{gautam2025memory_trans_main}. \textbf{c} and \textbf{d,} Homodyne interference plots with MW pulses either ON with varying phase or OFF. The drop in counts between MW ON and OFF data-point at $\pi/2$-phase depicts the contribution of transduced counts. \textbf{e} and \textbf{f,} Homodyne interference plots with MW pulses ON but with varying phase and 0-phase. The drop in counts around $\pi-$phase difference is the signature of destructive interference between the transduced photons and the local oscillator.}  \label{fig:Optical_Interference}
\end{figure*}

In addition to the data presented in Fig.~4~\cite{gautam2025memory_trans_main} of the main text, we present more interference measurements data in this section. Fig.~\ref{fig:Optical_Interference}(a and b), are data for interference experiments with sequential pair of MW pulses with the frequency of the later MW pulse varying at a constant phase difference $(\Delta \phi = \pi/2 \ \text{and} \ \pi)$ from the early MW. \par

Fig.~\ref{fig:Optical_Interference}(c to f) shows the results from Homodyne interference. In Fig.~\ref{fig:Optical_Interference}(c and d), the phase of MW1 and MW2 are varied when they are ON and fitted to a sinusoidal visibility function. The average photon counts are lower when the MW pulses are OFF. In Fig.~\ref{fig:Optical_Interference}(e and f)  MW1 and MW2 both are ON with the blue points taken with varying MW phases while the green points with constant MW phases in each plot. The counts are similar at 0-phase and stays flat when the MW phase is not changed, while it results in a sinusoidal pattern when the MW phase is varied. 

\section{More Data on Spectro-Temporal Multiplexing}

\begin{figure*}[h!]
    \centering  \includegraphics[width=0.75\linewidth]{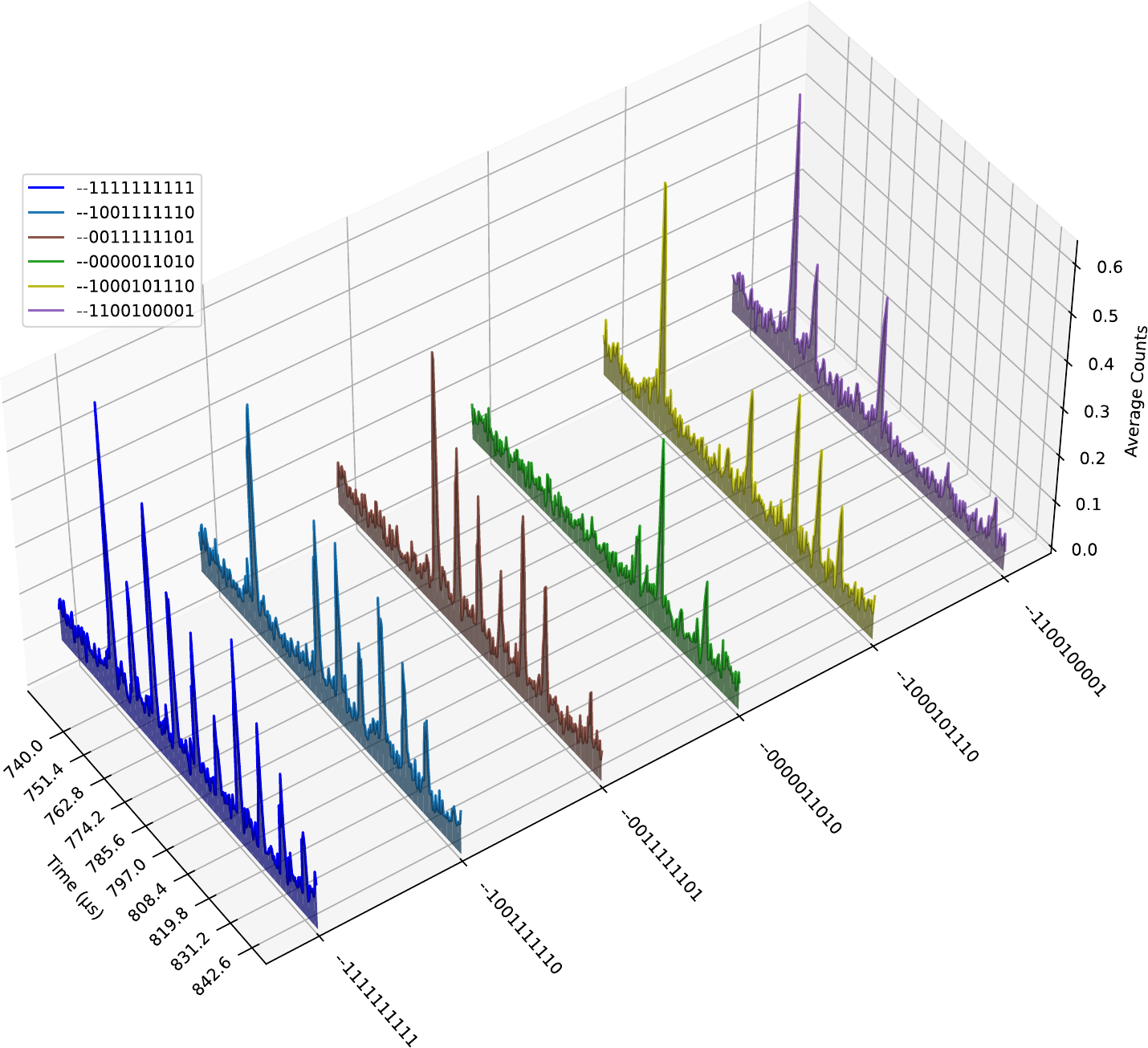}
    \caption{\textbf{Spectro-temporal multiplexing:} Stacked 3D plot of the data presented in the multiplexing section of the main text to clearly depict the counts for individual MW strings.}  \label{fig:Multiplexing_3D}
\end{figure*}

\begin{figure}[h!]
    \centering  \includegraphics[width=1\linewidth]{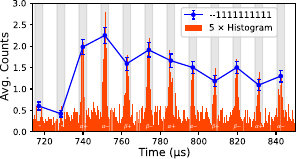}
    \caption{\textbf{Additional multiplexing trial:} Another trial of the multiplexing experiment showing similar level of counts from each cell for the first 4-modes. It decreases for the next set of 4-modes. Note that the second and 4th modes have higher counts than the first and third counts respectively. This is an effect of the AOM shift efficiency. This data is taken with a slightly different optical coupling through the AOM than the data presented in the main text.}  \label{fig:Multiplexing_repeat}
\end{figure}

\subsection{Microwave-Spectral Multiplexing}\label{MW_spectral_multiplex}
\begin{figure}[h!]
    \centering  \includegraphics[width=\linewidth]{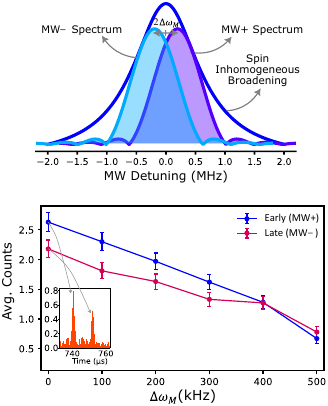}
    \caption{\textbf{Microwave-Spectral Multiplexing. Top,} Schematic of Fourier spectra of two MW pulses symmetrically detuned from the centre of the spin inhomogeneous broadening. \textbf{Bottom}, Sweep of the MW detuning with pump frequency fixed at the offset centre of the RAP spectrum. The inset shows the corresponding temporal histogram.}  \label{fig:MW_Multiplexing}
\end{figure}

By exploiting the spin inhomogeneous broadening, it is possible to transduce multiple MW pulses at distinct frequencies within the broadening profile. To ensure efficient absorption, the spectral bandwidth of each MW pulse must be confined within the extent of the spin inhomogeneous broadening. An illustration of this concept is shown in Fig.~\ref{fig:MW_Multiplexing} (Left), where the frequency spectra of two MW pulses are overlaid on the spin broadening profile. Since the MW pulses are weak in power, spectral overlap between them does not significantly affect their transduction efficiencies. We experimentally demonstrate MW spectral multiplexing by employing two-pairs of pulses with the same optical pump but distinct MW frequencies given by $\omega_M^\pm = \omega_{spin}\pm \Delta \omega_M$. The frequency separation $2\Delta \omega_M$ is varied within the spin inhomogeneous linewidth ($\Delta \omega_M \in [-0.5, 0.5]\,\rm{MHz}$). The experimental result is presented in Fig.~\ref{fig:MW_Multiplexing} (Right), where the transduction efficiency decreases with increasing detuning for both the pulses (early and late). The higher counts for early pulse are due to the single-frequency pump pulse, as discussed in main text sec.~II~E~1~\cite{gautam2025memory_trans_main}. \par 

%\FloatBarrier
\clearpage

%%%%%%%%%%%%%%%%%%%%%%%%%%%%%%%%%%%%%%%%%%%%%%%%%%%
% \bibliographystyle{unsrt}
\bibliography{main_refs}% Produces the bibliography via BibTeX.

% \bibliographystyle{apsrev4-2}
%%%%%%%%%%%%%%%%%%%%%%%%%%%%%%%%%%%%%%%%%%%%%%%%%%%%%%%%

\end{document}